\documentclass[lettersize,journal]{IEEEtran}
\usepackage{amsmath,amssymb,amsfonts}
\usepackage{array}
\usepackage[caption=false,font=normalsize,labelfont=sf,textfont=sf]{subfig}
\usepackage[toc,acronym]{glossaries}
\usepackage[inline]{enumitem}
\usepackage{algorithmic}
\usepackage{algorithm}
\usepackage{nicematrix}
\usepackage{booktabs}
\usepackage{orcidlink}
\usepackage{textcomp}
\usepackage{stfloats}
\usepackage{url}
\usepackage{verbatim}
\usepackage{graphicx}
\usepackage{balance}
\usepackage{nicematrix}
\usepackage{cite}
\usepackage{xcolor}
\usepackage{tikz}
\usetikzlibrary{positioning, shapes, arrows, calc}
\usepackage{pifont}

\newif\ifrevision
\revisionfalse
\ifrevision
  \newcommand{\hl}[1]{\textcolor{blue}{#1}}  
\else
  \newcommand{\hl}[1]{#1} 
\fi

\newcommand{\cmark}{\ding{51}}
\newcommand{\xmark}{\ding{55}}
\newcommand{\tableheader}{blue!30}
\newcommand{\tablerowcolor}{blue!5}

\definecolor{headercolor}{rgb}{0.84,1,1}
\definecolor{payloadcolor}{RGB}{204,229,255}
\definecolor{lightgray}{RGB}{211,211,211}

\newacronym{6lowpan}{6LoWPAN}{IPv6 over Low-Power Wireless Personal Area Networks}
\newacronym{ai}{AI}{Artificial Intelligence}
\newacronym{ap}{AP}{Access Point}
\newacronym{api}{API}{Application Program Interface}
\newacronym{ae}{AE}{Autoencoder}
\newacronym{automl}{AutoML}{Automated Machine Learning}
\newacronym{asn}{ASN}{Absolute Slot Number}
\newacronym{beassa}{BEA-SSA}{Bald Eagle Assisted SSA}
\newacronym{bs}{BS}{Base Station}
\newacronym{bfs}{BFS}{Breadth First Search}
\newacronym{blip}{BLIP}{Berkeley Low-power IP}
\newacronym{ch}{CH}{Cluster Head}
\newacronym{ctp}{CTP}{Collection Tree Protocol}
\newacronym{cnn}{CNN}{Convolutional Neural Networks}
\newacronym{dt}{DT}{Decision Tree}
\newacronym{dcnn}{DCNN}{Deep Convolutional Neural Network}
\newacronym{dl}{DL}{Deep Learning}
\newacronym{dqn}{DQN}{Deep Q-Learning}
\newacronym{dodag}{DODAG}{Destination Oriented Directed Acyclic Graph}
\newacronym{drl}{DRL}{Deep Reinforcement Learning}
\newacronym{eos}{EOS}{Embedded Operating System}
\newacronym{eb}{EB}{Enhanced Beacon}
\newacronym{ewma}{EWMA}{Exponential Weighted Moving Average}
\newacronym{fl}{FL}{Federated Learning}
\newacronym{fnd}{FND}{First Node Death}
\newacronym{fpga}{FPGA}{Field-Programmable Gate Array}
\newacronym{fsm}{FSM}{Finite State Machine}
\newacronym{gan}{GAN}{Generative Adversarial Network}
\newacronym{gprs}{GPRS}{General Packet Radio Service}
\newacronym{gps}{GPS}{Global Positioning System}
\newacronym{glpk}{GLPK}{GNU Linear Programming Kit}
\newacronym{hnd}{HND}{Half Node Death}
\newacronym{hrl}{HRL}{Hierarchical Reinforcement Learning}
\newacronym{hrltsch}{HRL-TSCH}{Hierarchical Reinforcement Learning-based Time Slotted Channel Hopping}
\newacronym{ia}{IA}{Intelligent Agent}
\newacronym{icmpv6}{ICMPv6}{Internet Control Message Protocol version 6}
\newacronym{ietf}{IETF}{Internet Engineering Task Force}
\newacronym{iot}{IoT}{Internet of Things}
\newacronym{iiot}{IIoT}{Industrial Internet of Things}
\newacronym{ids}{IDS}{Intrusion Detection System}
\newacronym{ip}{IP}{Internet Protocol}
\newacronym{ipv4}{IPv4}{Internet Protocol version 4}
\newacronym{ipv6}{IPv6}{Internet Protocol version 6}
\newacronym{knn}{k-NN}{K-Nearest Neighbour}
\newacronym{kpi}{KPI}{Key Performance Indicator}
\newacronym{lan}{LAN}{Local Area Network}
\newacronym{leach}{LEACH}{Low-Energy Adaptive Clustering Hierarchy}
\newacronym{leach-c}{LEACH-C}{Low-Energy Adaptive Clustering Hierarchy with Centralized Controller}
\newacronym{leach-rlc}{LEACH-RLC}{Low-Energy Adaptive Clustering Hierarchy with Reinforcement Learning-based Controller}
\newacronym{lt}{LT}{Lifetime}
\newacronym{lr}{LR}{Learning Rate}
\newacronym{wlan}{WLAN}{Wireless Local Area Network}
\newacronym{ler}{LER}{Listening Efficiency Ratio}
\newacronym{lora}{LoRa}{Long Range}
\newacronym{lorawan}{LoRaWAN}{Long Range Wide Area Network}
\newacronym{lowpan}{LoWPAN}{Low-Power Wireless Personal Area Networks}
\newacronym{lqi}{LQI}{Link Quality Indicator}
\newacronym{m2m}{M2M}{Machine-to-Machine}
\newacronym{mac}{MAC}{Media Access Control}
\newacronym{mdp}{MDP}{Markov Decision Process}
\newacronym{mems}{MEMS}{Micro-Electro-Mechanical Systems}
\newacronym{mcu}{MCU}{Microcontroller Unit}
\newacronym{milp}{MILP}{Mixed Integer Linear Programming}
\newacronym{ml}{ML}{Machine Learning}
\newacronym{mlsdwsn}{ML-SDWSN}{Machine Learning Software-Defined Wireless Sensor Network}
\newacronym{mst}{MST}{Minimum Spanning Tree}
\newacronym{na}{NA}{Neighbor Advertisement}
\newacronym{nc}{NC}{Network Configuration}
\newacronym{nes}{NES}{Networked Embedded System}
\newacronym{nd}{ND}{Neighbor Discovery}
\newacronym{nl}{NL}{Network Lifetime}
\newacronym{nn}{NN}{Neural Network}
\newacronym{nch}{NCH}{Non-Cluster Head}
\newacronym{os}{OS}{Operating System}
\newacronym{pdr}{PDR}{Packet Delivery Ratio}
\newacronym{plr}{PLR}{Packet Loss Rate}
\newacronym{pso}{PSO}{Particle Swarm Optimisation}
\newacronym{pca}{PCA}{Principal Component Analysis}
\newacronym{qos}{QoS}{Quality of Service}
\newacronym{ram}{RAM}{Random-Access Memory}
\newacronym{rdc}{RDC}{Radio Duty-Cycle}
\newacronym{rom}{ROM}{Read-Only Memory}
\newacronym{rnn}{RNN}{Recurrent Neural Network}
\newacronym{rl}{RL}{Reinforcement Learning}
\newacronym{rlasl}{RL-ASL}{Reinforcement Learning-based Adaptive Slot Listening}
\newacronym{rtt}{RTT}{Round-Trip Time}
\newacronym{rpl}{RPL}{Routing Protocol for Low-Power and Lossy Networks}
\newacronym{rssi}{RSSI}{Received Signal Strength Indicator}
\newacronym{stp}{STP}{Spanning Tree Protocol}
\newacronym{svm}{SVM}{Support Vector Machine}
\newacronym{sdn}{SDN}{Software-Defined Networking}
\newacronym{snr}{SNR}{Signal to Noise Ratio}
\newacronym{slip}{SLIP}{Serial Line Internet Protocol}
\newacronym{sdwsn}{SDWSN}{Software-Defined Wireless Sensor Network}
\newacronym{soh}{SOH}{State of Health}
\newacronym{sp}{SP}{Shortest Path}
\newacronym{tdma}{TDMA}{Time Division Multiple Access}
\newacronym{tl}{TL}{Transfer Learning}
\newacronym{tw}{TW}{Time Window}
\newacronym{2h}{\(2\text{-}\mathcal{H}\)}{Two-Handshake Synchronization Mechanism}
\newacronym{tlm}{TLM}{Traffic Load Minimisation}
\newacronym{tcp}{TCP}{Transmission Control Protocol}
\newacronym{tsch}{TSCH}{Time Slotted Channel Hopping}
\newacronym{udp}{UDP}{User Datagram Protocol}
\newacronym{uip}{$\mu$IP}{micro Internet Protocol}
\newacronym{uipv6}{$\mu$IPv6}{micro Internet Protocol version 6}
\newacronym{wam}{WAM}{Weighted Arithmetic Mean}
\newacronym{wban}{WBAN}{Wireless Body Area Network}
\newacronym{wpan}{WPAN}{Wireless Personal Area Network}
\newacronym{wsan}{WSAN}{Wireless Sensor and Actuator Network}
\newacronym{wsn}{WSN}{Wireless Sensor Network}

\begin{document}
\title{\acrshort{rlasl}: A Dynamic Listening Optimization for \acrshort{tsch} Networks Using \acrlong{rl}}
\author{F. Fernando~Jurado-Lasso\,\orcidlink{0000-0002-5005-781X}, \IEEEmembership{Member, IEEE},
  and J. F. Jurado\,\orcidlink{0000-0001-5193-8566}
  \thanks{Manuscript received June 9, 2025; revised xx, xx.}
  \thanks{F. Fernando Jurado-Lasso is an independent researcher, based in Cali, Colombia (e-mail: fdo.jurado@gmail.com).}
  \thanks{J. F. Jurado is with the Department of Basic Science, Faculty of Engineering and Administration, Universidad Nacional de Colombia Sede Palmira, Palmira 763531, Colombia (e-mail: jfjurado@unal.edu.co).}
}

\markboth{Journal of \LaTeX\ Class Files,~Vol.~18, No.~9, June~2025}%
{Jurado-Lasso \MakeLowercase{\textit{et al.}}: RL-ASL: A Dynamic Listening Optimization for TSCH Networks Using RL}

\maketitle

\begin{abstract}
  \acrfull{tsch} is a widely adopted \acrfull{mac} protocol within the IEEE 802.15.4e standard, designed to provide reliable and energy-efficient communication in \acrfull{iiot} networks.
However, state-of-the-art \acrshort{tsch} schedulers rely on static slot allocations, resulting in idle listening and unnecessary power consumption under dynamic traffic conditions.
This paper introduces \textbf{RL-ASL}, a reinforcement learning–driven adaptive listening framework that dynamically decides whether to activate or skip a scheduled listening slot based on real-time network conditions.
By integrating learning-based slot skipping with standard \acrshort{tsch} scheduling, RL-ASL reduces idle listening while preserving synchronization and delivery reliability.
Experimental results on the FIT IoT-LAB testbed and Cooja network simulator show that RL-ASL achieves up to \textbf{46\% lower power consumption} than baseline scheduling protocols, while maintaining \textbf{near-perfect reliability} and reducing average latency by up to \textbf{96\%} compared to PRIL-M.
Its link-based variant, RL-ASL-LB, further improves delay performance under high contention with similar energy efficiency.
Importantly, RL-ASL performs inference on constrained motes with negligible overhead, as model training is fully performed offline.
Overall, RL-ASL provides a practical, scalable, and energy-aware scheduling mechanism for next-generation low-power \acrshort{iiot} networks.
\end{abstract}


\section{Introduction}

\IEEEPARstart{T}{he} rapid evolution of \acrfullpl{nes} has revolutionized the Internet of Things (\acrshort{iot}), enabling seamless connectivity and intelligent decision-making across diverse applications.
\acrshort{iot} technologies are integral to critical domains such as industrial automation, healthcare monitoring, and smart grid systems~\cite{nguyen6GInternetThings2022,sooriInternetThingsSmart2023,jurado-lassoSurveyMachineLearning2022b}.
These systems require highly reliable, energy-efficient communication protocols to ensure sustained performance and scalability under varying conditions.

Among the communication technologies that support \acrshort{iot} applications, the \acrfull{tsch} protocol has emerged as a leading solution for robust and deterministic networking~\cite{watteyneUsingIEEE802154e2015a}.
By combining time synchronization and channel hopping, \acrshort{tsch} achieves resilience against interference and multipath fading, making it particularly well-suited for \acrfull{iiot} deployments.
Communication in \acrshort{tsch} is organized into synchronized time slots, with nodes adhering to predefined schedules that determine when to transmit, receive, or remain idle~\cite{urkeSurvey802154TSCH2021,zhang6TiSCHIIoTNetwork2024}.
This deterministic operation minimizes collisions and provides predictable performance, which is essential for industrial applications requiring high reliability and low latency.

Despite these advantages, \acrshort{tsch} networks suffer from a key inefficiency known as \textit{idle listening}—a condition where nodes keep their radios active during receive (Rx) slots without any actual packet transmission.
This behavior, often occurring in networks with sporadic or low traffic, leads to unnecessary energy expenditure.
In long-lived \acrshort{iot} deployments with battery-powered or energy-harvesting devices, mitigating idle listening is crucial to extending network lifetime and sustaining performance.

The proposed solution targets industrial and environmental monitoring systems where both energy efficiency and responsiveness are critical.
Examples include process supervision and anomaly detection in production plants, or safety and condition monitoring in warehouse and logistics infrastructures.
In such deployments, sensors must operate for years without maintenance, while still being capable of promptly reporting critical events such as temperature deviations, vibration anomalies, or air-quality alerts.
Unlike traditional environmental monitoring systems, these applications require a balanced energy-latency tradeoff: energy must be conserved to extend device lifetime, but latency cannot be excessively sacrificed without degrading responsiveness and system reliability.
Battery replacement is often costly or infeasible in these settings, and prolonged reporting delays may cause product degradation or safety risks.
Therefore, adaptive listening mechanisms that simultaneously achieve low power consumption and reduced latency are essential for next-generation \acrshort{iiot} systems.

Although TSCH schedules are deterministic, deciding whether a node should activate its radio in a given reception slot remains challenging due to uncertainty in effective transmission timing caused by retransmissions, traffic, etc. Static heuristics or explicit signaling often lead to conservative listening strategies with increased latency or protocol overhead.

To address this challenge, this paper proposes \acrfull{rlasl}, a \acrshort{rl}–based robust, deployment-time adaptive listening policy designed for environments where traffic patterns may vary within a known operational range.
Responsiveness is achieved through rapid, local runtime decisions based on a pre-trained policy, while energy efficiency is ensured by minimizing idle listening through learned slot-skipping strategies.
\acrshort{rlasl} dynamically decides whether a node should listen or skip a slot based on learned traffic and scheduling patterns, thereby reducing idle listening without compromising packet reliability.
Importantly, the proposed approach is agnostic to how TSCH schedules are generated. RL-ASL can operate on top of both static schedules and adaptive TSCH schedulers that dynamically add, remove, or relocate cells based on traffic demand or network conditions, as it relies only on locally observable slot-level information at runtime.
The learning phase is performed offline using extensive simulations that cover multiple representative traffic patterns, and the resulting policies are aggregated into a single generalized Q-table using \acrfull{fl}.
This generalized policy enables nodes to adapt their listening behavior at runtime based solely on locally observed state information, without requiring retraining or redeployment when traffic patterns vary within the trained regime.
Here, adaptivity refers to per-slot, runtime decision-making based on observed traffic dynamics, rather than online policy retraining or network-wide reconfiguration after deployment.
Scenarios involving highly unpredictable or adversarial traffic changes may require complementary mechanisms, such as centralized reconfiguration or online learning, which are outside the scope of this work.
Unlike prior work, \acrshort{rlasl} is fully implemented in \textsc{Contiki-NG}~\cite{oikonomouContikiNGOpenSource2022a} and evaluated on the FIT IoT-LAB~\cite{adjihFITIoTLABLarge2015a} testbed, ensuring both realism and reproducibility.
We also provide an independent implementation of PRIL-M~\cite{scanzioUltralowPowerGreen2024}, the closest state-of-the-art adaptive listening protocol.
Our experimental results demonstrate that policies trained on simple simulated topologies remain effective when deployed on larger and structurally different real-world networks, including relay nodes experiencing traffic loads not explicitly seen during training.
All our code and experimental artifacts are publicly released to foster further research in adaptive listening for \acrshort{tsch} networks.

\subsection{Contributions}

This paper makes the following key contributions:

\begin{enumerate}
    \item \textit{Proposing \acrshort{rlasl}:} We introduce a novel \acrshort{rl}–based adaptive listening protocol for \acrshort{tsch} networks that significantly reduces idle listening and improves energy efficiency.
    \item \textit{Formal Modeling and Constraints:} We formalize the listen-receive and listen-skip decision constraints that govern adaptive listening behavior, establishing a rigorous framework for protocol design.
    \item \textit{Generalized Offline Learning:} We design an offline training methodology based on diverse traffic patterns and \acrshort{fl}, yielding a single Q-table that generalizes across varying network conditions without requiring online retraining.
    \item \textit{Full Implementation in \textsc{Contiki-NG}:} We implement \acrshort{rlasl} entirely within the \textsc{Contiki-NG} operating system, ensuring compatibility with real-world \acrshort{tsch} deployments and open testbeds.
    \item \textit{Experimental Evaluation on FIT IoT-LAB:} We evaluate \acrshort{rlasl} on the FIT IoT-LAB platform, comparing its performance with state-of-the-art protocols—particularly PRIL-M—across diverse network topologies and traffic patterns.
    \item \textit{Open-Source Release:} We publicly release both our \acrshort{rlasl} and PRIL-M implementations, enabling reproducibility and future comparative research in adaptive listening mechanisms~\footnote{The code is available at \url{https://github.com/fdojurado/contiki-ng-rl-asl}}.
\end{enumerate}

The remainder of this paper is organized as follows:
Section~\ref{sec:related_work} reviews related work on energy-efficient communication in \acrshort{tsch} networks.
Section~\ref{sec:network_model} presents the network model and discusses idle listening inefficiencies.
Section~\ref{sec:rl_asl} details the design of the \acrshort{rlasl} protocol.
Section~\ref{sec:implementation} describes implementation details and the experimental setup.
Section~\ref{sec:performance_metrics} outlines the performance metrics and baseline protocols used for evaluation.
Section~\ref{sec:results} presents and discusses the experimental results.
Finally, Section~\ref{sec:conclusion} concludes the paper and discusses future research directions.

\section{Related Work}
\label{sec:related_work}

Energy-efficient and reliable communication remains a core challenge in low-power wireless networks, especially in \acrshort{iiot} settings where long lifetimes and predictable performance are essential~\cite{rathoreEnablingFaultTolerance2022}.
The \acrshort{tsch} protocol has become a leading solution for such systems thanks to its deterministic time-slotting and channel-hopping mechanisms~\cite{taboucheTrafficAwareReliableScheduling2023b}.
However, \acrshort{tsch} networks still suffer from energy waste due to \textit{idle listening} and limited adaptability to dynamic traffic conditions.

\subsection{Autonomous and Centralized Scheduling Mechanisms}
Early research in \acrshort{tsch} scheduling focused on improving slot allocation to balance reliability and energy efficiency.
Orchestra~\cite{duquennoyOrchestraRobustMesh2015a} pioneered autonomous scheduling by letting each node independently assign its transmission and reception cells based on network topology.
This approach reduced coordination overhead and improved robustness but did not explicitly address the problem of idle listening, as nodes kept their radios active during all assigned Rx slots regardless of actual traffic.

Building upon this, Kim \textit{et al.} proposed ALICE~\cite{kimALICEAutonomousLinkbased2019}, a link-based scheduling protocol where each node allocates cells dynamically according to its neighbors’ traffic direction and updates schedules every slotframe.
ALICE improved upon Orchestra in terms of throughput and energy efficiency, yet nodes still incurred energy losses when listening to unused Rx slots.
Similarly, centralized and heuristic-based schedulers—such as those by Khan \textit{et al.}~\cite{ojoEnergyEfficientCentralized2017} and Papadopoulos \textit{et al.}~\cite{papadopoulosGuardTimeOptimisation2017}—achieved gains through global coordination or guard-time optimization but at the expense of scalability and autonomy.

More recent adaptive schedulers such as TESLA~\cite{jeongTESLATrafficawareElastic2019} and OST~\cite{jeongOSTOndemandTSCH2020} introduced elastic slotframe management, dynamically adjusting the slotframe length and cell allocation based on observed traffic intensity.
These methods improve throughput and energy efficiency under fluctuating loads but still require the radio to remain active during allocated Rx slots, leaving idle listening largely unaddressed.

\subsection{Recent Adaptive and Traffic-Aware Scheduling}
Recent advances have produced more sophisticated adaptive schedulers that monitor network traffic and topology in real time.
For example, A3~\cite{kimA3AdaptiveAutonomous2021} employs receiver-side traffic estimation to autonomously adjust the number of Tx/Rx cells in the slotframe, achieving notable adaptability and reduced congestion under dynamic traffic.
DT-SF~\cite{tavallaieDesignOptimizationTrafficAware2021} follows a similar principle by adjusting slot allocations according to traffic demands and link reliability metrics.
Similarly, TA-RPL~\cite{haTrafficAware6TiSCHRouting2022} extends adaptability beyond the MAC layer by integrating traffic-aware routing with TSCH scheduling to optimize end-to-end network performance.
It leverages the number of allocated transmission cells as an indicator of both link quality and traffic intensity, enabling routing decisions that balance load and improve bandwidth utilization across the network.
While these methods substantially improve scheduling agility, they primarily operate at the slotframe or routing level and do not consider fine-grained energy optimization within allocated Rx slots.
As a result, even adaptive schedulers may suffer from residual idle listening when nodes expect but do not receive traffic.

\subsection{Learning-Based Scheduling Approaches}
Reinforcement learning (RL) has recently emerged as a powerful tool for adaptive scheduling in constrained IoT environments.
Pratama \textit{et al.} proposed RL-SF~\cite{pratamaRLSFReinforcementLearning2022} and LLQL-SF~\cite{pratamaLowLatencyQLearningBasedDistributed2024}, which use Q-learning to optimize cell allocations and minimize latency across distributed nodes.
Similarly, ELISE~\cite{jurado-lassoELISEReinforcementLearning2024} and HRL-\acrshort{tsch}~\cite{jurado-lassoHRLTSCHHierarchicalReinforcement2024} apply hierarchical or model-based RL to adapt schedules to changing traffic patterns while maintaining deterministic communication.
Although these studies demonstrate that RL can outperform heuristic approaches in adapting to dynamic conditions, they focus on slotframe configuration or global scheduling optimization, not on reducing idle listening at the link level.

\subsection{Idle Listening Reduction Mechanisms}
Several mechanisms have been proposed specifically to mitigate idle listening in \acrshort{tsch} networks.
Nsabagwa \textit{et al.}~\cite{nsabagwaMinimalIdlelistenCentralized2018} formulated the scheduling problem as a constraint satisfaction problem (CSP) to minimize idle slots in centralized networks, achieving reduced delay and energy consumption but limited scalability.
PRIL-F~\cite{scanzioEnergySavingTSCH2020} and its successor PRIL~\cite{scanzioEnhancedEnergysavingMechanisms2023} introduced proactive radio-sleep coordination based on the \acrshort{asn}, allowing nodes to skip idle Rx slots.
More recently, PRIL-M~\cite{scanzioUltralowPowerGreen2024} improved robustness to acknowledgment loss and synchronization errors, demonstrating significant energy gains under periodic traffic.
However, PRIL-based methods rely on deterministic traffic patterns and shared synchronization, making them less effective in dynamic or bursty environments.

Kalita \textit{et al.} proposed OASA~\cite{kalitaOntheFlyAutonomousSlot2024}, an on-the-fly adaptive scheduler that adjusts slot allocations according to instantaneous traffic conditions, and TACTILE~\cite{kalitaAutonomousAllocationScheduling2021}, which distributes slots across multiple channels to reduce desynchronization.
While these mechanisms reduce energy consumption by adapting slot allocations, their heuristic nature limits long-term adaptability in non-stationary traffic environments.

Notably, PRIL and its variants constitute the existing body of work that explicitly targets adaptive unicast idle listening reduction in TSCH networks. Other recent TSCH proposals primarily address scheduling, routing, or slotframe elasticity and are therefore complementary rather than directly comparable to listening control mechanisms such as RL-ASL.

\subsection{Positioning of RL-ASL}
Despite substantial progress, no prior work directly addresses \textit{unicast idle listening} in distributed \acrshort{tsch} networks using a learning-based approach.
Existing adaptive and RL-based schedulers optimize slot allocation or slotframe parameters but leave per-slot radio activation unmanaged.
\acrshort{rlasl} fills this gap through a lightweight, distributed RL agent that learns per-link traffic patterns and decides whether to listen or skip each Rx slot.
Unlike rule-based or centralized schemes, it autonomously adapts to stochastic traffic while maintaining near-perfect reliability.
Operating at the granularity of individual receive opportunities, RL-ASL complements existing schedulers by enhancing energy efficiency without requiring global coordination or traffic predictability.

In summary, previous studies focus on slotframe- or network-level adaptation, leaving per-slot energy management largely unexplored.
RL-ASL bridges this gap by learning fine-grained listening behaviors that adapt to dynamic traffic, advancing link-level energy optimization in TSCH networks.

\section{Network Model and Idle Listening Inefficiency}
\label{sec:network_model}

This section presents the network model adopted in this work and defines the concept of idle listening inefficiency in \acrshort{tsch}-based networks.

\subsection{Network Model}

We consider a \acrshort{tsch} network composed of \( N \) nodes operating in a synchronized \textit{slotframe} of \( L \) time slots, indexed from \( 0 \) to \( L-1 \), with each slot of duration \( \tau \) seconds. The network time is represented by the \acrfull{asn}, which provides a global synchronization reference for slotframe alignment and coordinated transmissions.

Each node \( n \) listens for incoming unicast transmissions during one or more designated reception slots per slotframe, denoted by the set \( \mathcal{L}_n^k \subseteq \{0, 1, \dots, L-1\} \) in slotframe \( k \).

Communication occurs between node \( n \) and its set of neighbors \( \mathcal{M}_n \). The total number of listening activations for node \( n \) during an observation window of \( K \) consecutive slotframes is given by
\[
\sum_{k=1}^{K} |\mathcal{L}_n^k|,
\]
which represents the cumulative number of reception opportunities across time.

Nodes are categorized according to their functional role in the data collection process:

\subsubsection{Node Classification}
\begin{itemize}
    \item \textbf{Root node} (\(\mathcal{G}\)): Collects data from all other nodes.
    \item \textbf{Relay nodes} (\(\mathcal{Y}\)): Forward data toward the root.
    \item \textbf{Leaf nodes} (\(\mathcal{Z}\)): Generate and transmit data to relay or root nodes.
\end{itemize}

\subsubsection{Packet Transmission and Reception}
Packets transmitted from node \( n \) to node \( m \) are denoted as \( \psi_{n \to m} \in \Psi \), while successfully received packets or those detected through control coordination (e.g., broadcast notifications) are represented as \( \omega_{m \to n} \in \Omega \). Each transmission or reception event occurs within a specific slot \( l \) of slotframe \( k \).

\subsubsection{Idle Listening and Energy Waste}

Idle listening occurs when a node keeps its radio active in a reception slot but no packet is transmitted to it. This leads to unnecessary energy expenditure without contributing to data reception.

We distinguish two cases:
\begin{enumerate*}[label=(\roman*)]
    \item \textbf{Unnecessary listening:} The node is awake but no transmission is directed to it.
    \item \textbf{Necessary listening:} A transmission is attempted, even if it fails; this is not considered idle.
\end{enumerate*}

Formally, we define the idle listening indicator for node \( n \) in slotframe \( k \), slot \( l \), as:
\begin{equation}
    \label{eq:delta}
    \delta_n(k, l) =
    \begin{cases}
        1 & \text{if } l \in \mathcal{L}_n^k \text{ and } \nexists \psi_{m \to n}^{k,l} \in \Psi,\\
        0 & \text{otherwise.}
    \end{cases}
\end{equation}
This definition ensures that only truly idle listening instances are counted as wasted energy, consistent with \acrshort{tsch} operation principles.

\subsubsection{Traffic Generation Model}
Each node generates packets at random time intervals \( \Delta T_n \), modeled as
\(
    \Delta T_n \sim \mathcal{N}(\mu_n, \sigma_n^2),
\)
where \( \mu_n \) and \( \sigma_n \) represent the mean and standard deviation of packet generation intervals, respectively. The distribution is truncated at zero to ensure positive intervals. This model reflects practical low-power wireless behavior where jitter is intentionally introduced to mitigate collisions and improve channel fairness, as commonly implemented in operating systems such as \textsc{Contiki-NG}~\cite{oikonomouContikiNGOpenSource2022a} and \textsc{RIOT}~\cite{baccelliRIOTOSOS2013}.

\subsubsection{Network Topology and Communication Range}
Nodes are placed in a two-dimensional plane, with each node \( n \) located at coordinates \( \rho_n = (x_n, y_n) \), where
\(
x_n, y_n \in [x_{\text{min}}, x_{\text{max}}] \times [y_{\text{min}}, y_{\text{max}}].
\)
The Euclidean distance between two nodes \( n \) and \( m \) is
\(
d_{n,m} = \sqrt{(x_n - x_m)^2 + (y_n - y_m)^2}.
\)
Successful communication occurs when \( d_{n,m} \leq d_{\text{max}} \), where \( d_{\text{max}} \) denotes the maximum communication range.

\subsubsection{Slotframe and Channel Management}
We adopt the \textit{Orchestra} scheduling framework to coordinate transmissions across multiple slotframes:
\begin{enumerate*}[label=(\roman*)]
    \item time synchronization (beacon) slotframe,
    \item broadcast slotframe, and
    \item unicast slotframe.
\end{enumerate*}
Our focus is on the \textit{unicast slotframe}, where receiver nodes listen in a single scheduled Rx slot per slotframe (\( |\mathcal{L}_n^k| = 1 \)).  
The network uses \( C \) channel offsets, denoted by \( \mathcal{C} = \{0, 1, \dots, C-1\} \), mapped to physical frequencies through the \acrshort{tsch} hopping sequence to mitigate interference and enable spatial reuse.

\begin{figure}[!t]
    \centerline{\includegraphics[width=1\columnwidth]{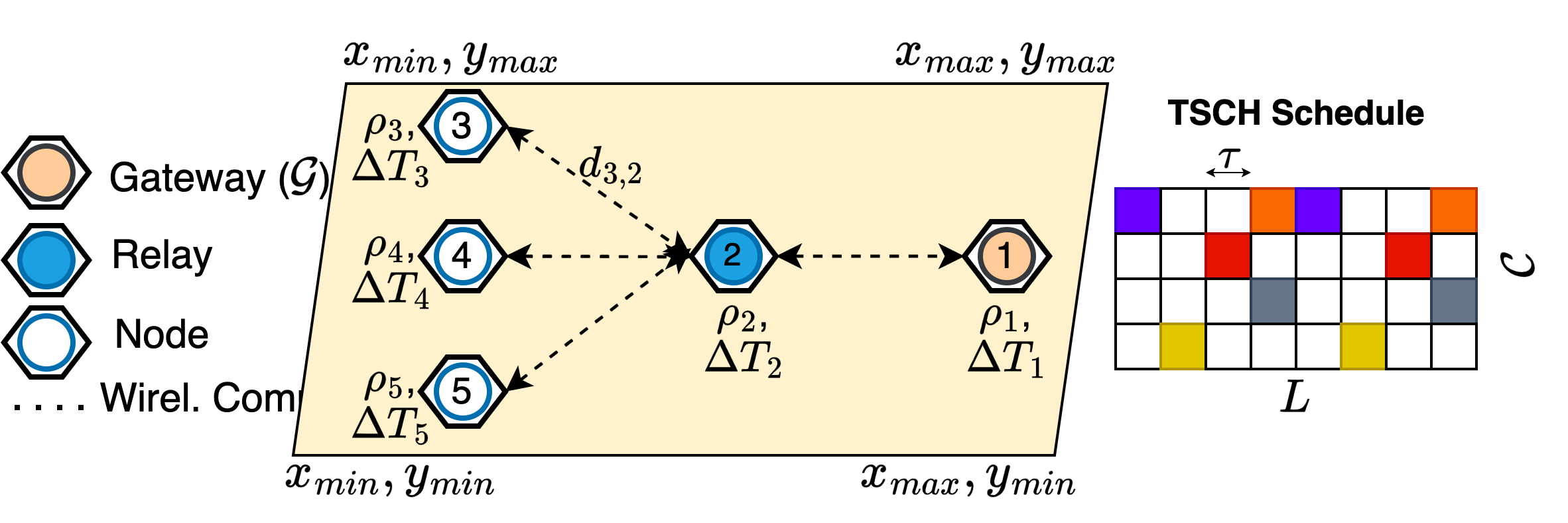}}
    \caption{Network model for a \acrshort{tsch} network with five nodes.}
    \label{fig:tsch_network_model}
\end{figure}

\section{RL-ASL: \acrshort{rl} for adaptive listening}
\label{sec:rl_asl}

The objective of RL-ASL is to minimize the overall energy cost associated with idle listening while maintaining high delivery reliability—a balance that can be viewed as minimizing a total energy cost \(J_\mathrm{total}\) under reliability constraints. RL-ASL achieves this adaptively through local Q-learning at each receiver, without global coordination.

RL-ASL is a fully distributed, receiver-side policy that uses tabular Q-learning to decide, at every scheduled unicast Rx slot, whether a node should \emph{listen} or \emph{skip} that slot. The design couples a compact per-neighbor temporal model (\acrfull{ewma} inter-arrival and variance) with a feature-engineered aggregated state, a probabilistic transmission model (distance-to-nearest Gaussian), and an expected-reward computation that trades energy savings against missed receptions. Nodes learn an \(\varepsilon\)-greedy tabular policy that is updated online (training mode) or loaded as a frozen table (evaluation mode).

\subsection{Neighbor model and notation}
\label{sec:neighbor_model}

For node \(n\) and each neighbor \(m \in \mathcal{M}_n\) we maintain a compact statistical descriptor
\[
  \Theta_{n,m} \;=\; \bigl(\, \mathrm{asn}_{n,m}^{\mathrm{last}},\; \hat{\mu}_{n,m},\; \hat{\sigma}_{n,m}^2,\; \mathrm{asn}_{n,m}^{\mathrm{exp}},\; \kappa_{n,m}\,\bigr),
\]
where:
\begin{itemize}
  \item \(\mathrm{asn}_{n,m}^{\mathrm{last}}\) is the \acrshort{asn} at which \(n\) last heard \(m\);
  \item \(\hat{\mu}_{n,m}\) is an EWMA estimate of the inter-arrival (period) between receptions from \(m\) (in slots);
  \item \(\hat{\sigma}_{n,m}^2\) is an EWMA estimate of the variance of those inter-arrivals;
  \item \(\mathrm{asn}_{n,m}^{\mathrm{exp}} = \mathrm{asn}_{n,m}^{\mathrm{last}} + \hat{\mu}_{n,m}\) is the next expected \acrshort{asn} for \(m\);
  \item \(\kappa_{n,m}\in\mathbb{Z}_{\ge0}\) counts consecutive predicted-but-missed receptions (predicted skips).
\end{itemize}

When node \(n\) observes a reception from \(m\) at \acrshort{asn} \(t_i\), the inter-arrival
\(\Delta_{n,m}(t_i)=\mathrm{asn}_{n,m}(t_i)-\mathrm{asn}_{n,m}(t_{i-1})\) is used to update the EWMA estimates:
\[
  \hat{\mu}_{n,m} \leftarrow (1-\lambda)\hat{\mu}_{n,m} + \lambda\,\Delta_{n,m}(t_i),
\]
\[
  \hat{\sigma}_{n,m}^2 \leftarrow (1-\lambda)\hat{\sigma}_{n,m}^2 + \lambda\big(\Delta_{n,m}(t_i)-\hat{\mu}_{n,m}\big)^2,
\]
with smoothing coefficient \(\lambda\in(0,1)\).  On a detected missed event we advance the expectation and increment \(\kappa_{n,m}\):
\[
  \mathrm{asn}_{n,m}^{\mathrm{exp}} \leftarrow \mathrm{asn}_{n,m}^{\mathrm{exp}} + \hat{\mu}_{n,m},\qquad
  \kappa_{n,m} \leftarrow \kappa_{n,m} + 1.
\]

These per-neighbor statistics are the primitives used by the probability model and the state encoding below.

\subsection{Probability-of-transmission model (distance-to-nearest)}
\label{sec:probability_model}

At current \acrshort{asn} \(a\) node \(n\) estimates the probability that \emph{at least one} child \(m\in\mathcal{M}_n\) will transmit in that slot. Let
\[
  \Delta_{n,m}(a) \;=\; a - \max\!\big(\mathrm{asn}_{n,m}^{\mathrm{last}},\,\mathrm{asn}_{n,m}^{\mathrm{exp}}\big).
\]
Denote \(\hat{\mu}_{n,m}\) and \(\hat{\sigma}_{n,m}\) the EWMA mean and standard deviation (square root of \(\hat{\sigma}_{n,m}^2\)) of inter-arrivals. To ensure numerical robustness we clamp \(\hat{\sigma}_{n,m}\) as:
\[
  \hat{\sigma}_{n,m} \leftarrow
  \min\!\big( \alpha\,\hat{\mu}_{n,m},\; \max(\sigma_{\min},\; \beta\,\hat{\mu}_{n,m},\; \hat{\sigma}_{n,m})\big),
\]
with configuration constants \(\alpha,\beta\in(0,1)\) and \(\sigma_{\min}>0\).

Define the \emph{phase} and \emph{distance-to-nearest}:
\begin{align*}
  \phi_{n,m}(a) & = \Delta_{n,m}(a)\bmod \hat{\mu}_{n,m},                           \\
  d_{n,m}(a)    & = \min\!\big(\phi_{n,m}(a),\; \hat{\mu}_{n,m}-\phi_{n,m}(a)\big).
\end{align*}
We model a per-neighbor instantaneous transmission likelihood as a Gaussian kernel on the distance-to-nearest:
\begin{equation}
  \label{eq:neighbor_prob_final}
  p_{n,m}(a) \;=\; \exp\!\Big(-\tfrac{1}{2}\frac{d_{n,m}^2(a)}{\hat{\sigma}_{n,m}^2}\Big),
  \qquad p_{n,m}(a)\in[0,1].
\end{equation}
Assuming conditional independence across neighbors, the probability that \emph{no} child transmits is
\(\prod_{m\in\mathcal{M}_n}(1-p_{n,m}(a))\), so the aggregated probability of at least one transmission is
\begin{equation}
  \label{eq:p_any_final}
  p_n(a) \;=\; 1 - \prod_{m\in\mathcal{M}_n}\!\big(1-p_{n,m}(a)\big),
\end{equation}
which the implementation clamps to \(p_n(a)\in[\varepsilon_p,\,1-\varepsilon_p]\) with \(\varepsilon_p=10^{-3}\) (approx.) to avoid degenerate expectations.

This distance-to-nearest model produces a temporal ``bump'' slightly before and after expected instants, providing permissive behavior under timing uncertainty.

\subsection{State encoding (feature engineering and aggregation)}
\label{sec:state_encoding}

To keep the Q-table compact, RL-ASL maps neighborhood statistics to a small discrete state \(s\in\mathcal{S}=\{0,\dots,S-1\}\) via feature engineering and mixed-radix encoding.

\paragraph{Per-neighbor bins.}
For node \(n\) with neighbors \(\mathcal{M}_n=\{m_1,\dots,m_{M_n}\}\), compute for each neighbor
\[
  b_i \;=\; \operatorname{bin}_B\!\Big(\frac{\Delta_{n,m_i}(a)}{\hat{\mu}_{n,m_i}}\Big),
  \quad i=1,\dots,M_n,
\]
where \(\operatorname{bin}_B(\cdot)\) maps the normalized elapsed inter-arrival to one of \(B\) discrete bins.

\paragraph{Neighborhood aggregates.}
From \(\{b_i\}\) and \(\{d_{n,m_i}(a)\}\) compute:
\begin{align*}
  \overline{b}       & = \operatorname{round}\!\Big(\frac{1}{M_n}\sum_{i=1}^{M_n} b_i\Big),                              \\
  c_{\mathrm{short}} & = \sum_{i=1}^{M_n}\mathbb{I}(b_i < b_{\mathrm{th}}),                                              \\
  d_{\min}(a)        & = \min_{m\in\mathcal{M}_n} d_{n,m}(a),\quad
  d_{\min}^{\mathrm{bin}} = \operatorname{bin}_D\big(d_{\min}(a)\big),                                                   \\
  c_{\mathrm{near}}  & = \min\!\Big(C_{\max},\,\sum_{m\in\mathcal{M}_n}\mathbb{I}(d_{n,m}(a)\le\hat{\sigma}_{n,m})\Big).
\end{align*}
Here \(b_{\mathrm{th}}\), \(C_{\max}\) are configuration constants, and \(\operatorname{bin}_D(\cdot)\) maps the minimum distance to one of \(D\) discrete bins.

\paragraph{Mixed-radix encoding.}
We form the aggregated state index via a bijection
\[
  s = f_{\mathrm{enc}}\!\big(\overline{b},\,c_{\mathrm{short}},\,d_{\min}^{\mathrm{bin}},\,c_{\mathrm{near}}\big),
\]
where \(S = B\cdot C_1\cdot D\cdot C_2\) (with \(C_1,C_2\) the discrete ranges of \(c_{\mathrm{short}},c_{\mathrm{near}}\)). All components are clipped to their declared ranges so \(s\) is guaranteed to satisfy \(0\le s < S\). This compact integer index is the input to the tabular Q-learner.

\subsection{Action space}
\label{sec:action_space}

At each scheduled unicast Rx slot (decision epoch) node \(i\) selects
\[
  a_t \in \mathcal{A} = \{a^{(0)},a^{(1)}\} = \{\texttt{SKIP RX},\ \texttt{DO NOT SKIP RX}\}.
\]
The radio activation indicator \(\xi_{i,t}\in\{0,1\}\) is determined by the chosen action:
\[
  \xi_{i,t} = \begin{cases}
    0 & \text{if } a_t = \texttt{SKIP RX},        \\
    1 & \text{if } a_t = \texttt{DO NOT SKIP RX}.
  \end{cases}
\]

\subsection{Reward shaping and expected-reward computation}
\label{sec:reward_shaping}
To explicitly account for the cost of missed transmissions, RL-ASL uses an expected-reward formulation that penalizes skipping reception slots proportionally to the estimated likelihood of an incoming transmission.
Let the configured reward/penalty parameter set be
\[
  \mathcal{R}=\{R_{\mathrm{succ}}>0,\; R_{\mathrm{skip}}>0,\; C_{\mathrm{idle}}<0,\; C_{\mathrm{miss}}<0\},
\]
interpreted as success reward, skip reward, idle-listen cost and miss penalty respectively.

Given the aggregated transmission probability \(p_{i,t}\equiv p_i(a_t)\) computed by Eq.~\eqref{eq:p_any_final}, the \emph{expected immediate reward} for node \(i\) taking action \(a_t\) is:
\begin{equation}
  \label{eq:expected_reward_final}
  \mathbb{E}[r_{i,t}\mid a_t] =
  \begin{cases}
    p_n(a_t)\,C_{\mathrm{miss}} + (1-p_n(a_t))\,R_{\mathrm{skip}}, \\[6pt]
    p_n(a_t)\,R_{\mathrm{succ}} + (1-p_n(a_t))\,C_{\mathrm{idle}}, \\
  \end{cases}
\end{equation}

The implementation uses this expectation as follows:
\begin{itemize}
  \item If \(a_t=\texttt{SKIP RX}\) there is no observation in the current slot: the Q-update uses \(\mathbb{E}[r_{i,t}\mid a_t]\) immediately.
  \item If \(a_t=\texttt{DO NOT SKIP RX}\) the agent listens; upon slot completion the actual outcome is observed. If a packet arrived then \(r_{i,t}=R_{\mathrm{succ}}\); otherwise the recomputed \(p_{i,t}\) is used and the update target is \(\mathbb{E}[r_{i,t}\mid a_t]\).
\end{itemize}

Two heuristic adjustments improve stability:
\begin{enumerate}
  \item \textbf{Missed-neighbor penalty:} if any neighbor \(j\) satisfies \(\Delta_{n,j}(a) \ge \hat{\mu}_{n,j} + \zeta_{\mathrm{miss}}\hat{\sigma}_{n,j}\)
  \item \textbf{Near-transmit penalty:} if multiple neighbors are near their expected transmit (within one \(\hat{\sigma}\)), the expected reward for skipping is decreased proportionally to the near count (implementation uses a small logged penalty to bias learning away from skipping).
\end{enumerate}

\subsection{Tabular Q-learning and learning dynamics}
\label{sec:q_learning}

Each node \(i\in\mathcal{N}\) maintains a local tabular action–value function
\[
  Q_i:\mathcal{S}\times\mathcal{A}\to\mathbb{R},
\]
parameterized by learning rate \(\alpha\in(0,1]\), discount factor \(\gamma\in(0,1)\), and exploration rate \(\varepsilon_e\).
At each decision epoch \(t\), after observing state \(s_{i,t}\), selecting action \(a_{i,t}\), and receiving or estimating the immediate reward \(r_{i,t}\),
the Q-table is updated according to the standard Q-learning rule:
\begin{equation}
  \label{eq:q_update_final}
  \begin{split}
    Q_i(s_{i,t},a_{i,t}) &\leftarrow Q_i(s_{i,t},a_{i,t}) \\
    &\quad + \alpha\Bigl[r_{i,t}
    + \gamma\max_{a'\in\mathcal{A}} Q_i(s_{i,t+1},a')
    - Q_i(s_{i,t},a_{i,t})\Bigr].
  \end{split}
\end{equation}

Action selection follows an \(\varepsilon\)-greedy policy:
\[
  a_{i,t} =
  \begin{cases}
    \arg\max_{a\in\mathcal{A}} Q_i(s_{i,t},a), & \text{w.p. } 1-\varepsilon_e, \\[4pt]
    \text{Uniform}( \mathcal{A} ),             & \text{w.p. } \varepsilon_e,
  \end{cases}
\]
where \(\varepsilon_e\) decays multiplicatively per episode as
\(\varepsilon_{e+1} = \max(\varepsilon_{\min},\,\eta_\varepsilon \varepsilon_e)\).
Each node maintains episode statistics
\(
G_e=\sum_{t=0}^{T_e-1} r_{i,t}
\)
and a rolling average \(\overline{G}_{\mathrm{roll}}\)
over a sliding window of \(W\) recent episodes.
Whenever \(\overline{G}_{\mathrm{roll}}\) exceeds its previous maximum,
the table may be checkpointed for debugging or persistence.
During training, Q-values are updated online; in evaluation mode,
the Q-table is frozen and read-only.

\subsection{Runtime decision and learning loop}
\label{sec:runtime_flow}

Algorithm~\ref{alg:rlasl} summarizes the per-node logic of RL-ASL.
At each scheduled unicast reception slot of node \(i\) at \acrshort{asn} \(a\):

\begin{enumerate*}[label=(\roman*)]
  \item \textit{Feature extraction:} the node computes per-neighbor features \(\{b_{n,m}(a)\}\), aggregates them into \((\overline{b}, c_{\mathrm{short}}, d_{\min}^{\mathrm{bin}}, c_{\mathrm{near}})\), and encodes the resulting state \(s_{i,t}=f_{\mathrm{enc}}(\cdot)\) as in Section~\ref{sec:state_encoding}.
  \item \textit{Action selection:} an action \(a_{i,t}\in\mathcal{A}\) is drawn from the \(\varepsilon\)-greedy policy on \(Q_i(s_{i,t},\cdot)\).
  \item \textit{Decision execution:} - If \(a_{i,t}=\texttt{SKIP RX}\), the node keeps the radio off,
        estimates \(p_{i,t}\) by Eq.~\eqref{eq:p_any_final},
        computes the expected reward \(\mathbb{E}[r_{i,t}\mid a_{i,t}]\)
        by Eq.~\eqref{eq:expected_reward_final},
        and applies the Q-update~\eqref{eq:q_update_final} immediately
        (using \(s_{i,t+1}=s_{i,t}\)).
        - If \(a_{i,t}=\texttt{DO NOT SKIP RX}\),
        the node listens; upon slot completion it observes whether a packet
        was received and sets
        \( r_{i,t} = R_{\mathrm{succ}} \) if packet received; otherwise \( r_{i,t} = \mathbb{E}[r_{i,t}\mid a_{i,t}] \),
        before performing the same Q-update.
  \item \textit{Bookkeeping:} increment the step counter, update
        \(G_e\) and \(\overline{G}_{\mathrm{roll}}\), and decay \(\varepsilon_e\)
        when the episode ends.
\end{enumerate*}

\begin{algorithm}[htbp]
  \scriptsize
  \caption{RL-ASL: Per-node online decision and learning loop}
  \label{alg:rlasl}
  \begin{algorithmic}[1]
    \FOR{each scheduled Rx slot of node \(i\) at \acrshort{asn} \(a\)}
    \IF{not associated or \(|\mathcal{M}_i|=0\)} \STATE listen (\(\xi_{i,t}\!\leftarrow\!1\)); Continue \ENDIF
    \STATE extract features and encode state \(s_{i,t}=f_{\mathrm{enc}}(\cdot)\)
    \STATE select \(a_{i,t}\) via \(\varepsilon\)-greedy on \(Q_i(s_{i,t},\cdot)\)
    \IF{\(a_{i,t}=\texttt{SKIP RX}\)}
    \STATE compute \(p_{i,t}\) via Eq.~\eqref{eq:p_any_final}
    \STATE compute expected reward \(\bar r_{i,t}\) via Eq.~\eqref{eq:expected_reward_final}
    \STATE update \(Q_i(s_{i,t},a_{i,t})\) using \(\bar r_{i,t}\)
    \ELSE
    \STATE activate radio (\(\xi_{i,t}\!\leftarrow\!1\))
    \IF{packet received} \STATE \(r_{i,t}\!\leftarrow\!R_{\mathrm{succ}}\)
    \ELSE \STATE recompute \(p_{i,t}\); \(r_{i,t}\!\leftarrow\!\mathbb{E}[r_{i,t}\mid a_{i,t}]\)
    \ENDIF
    \STATE update \(Q_i(s_{i,t},a_{i,t})\) using \(r_{i,t}\)
    \ENDIF
    \STATE bookkeeping (update counters, episode return, decay \(\varepsilon_e\))
    \ENDFOR
  \end{algorithmic}
\end{algorithm}

\subsection{Complexity and robustness considerations}

All computations are local and lightweight:
the Q-table size is \(|\mathcal{S}|\cdot|\mathcal{A}|\) entries,
and per-neighbor statistics require a few floating-point scalars.
The only nonlinear operations are \(\exp(\cdot)\), \(\sqrt{\cdot}\),
and modular arithmetic for the phase calculation.
Numerical stability is ensured through bounded counters,
clamping of probabilities \(p_{i,t}\in[\varepsilon_p,1-\varepsilon_p]\),
and limiting \(\hat{\sigma}_{n,m}\) as defined in
Section~\ref{sec:probability_model}.

\subsection{Discussion on Mobility Support}

Mobility-induced topology changes, such as those triggered by upper-layer routing protocols (e.g., RPL), can be naturally accommodated by RL-ASL through its \acrshort{2h} (see Section~\ref{sec:implementation_details}).
Routing updates in RPL-based networks may take several minutes~\cite{vallatiImprovingNetworkFormation2019}; during this period, affected nodes revert to standard \acrshort{tsch} behavior, listening to all scheduled Rx slots to maintain network connectivity and discover new neighbors.
Once routing converges and a new parent is selected, the node reinitiates the 2H process to synchronize with the new receiver.
After successful coordination, RL-ASL resumes normal operation, allowing the \acrshort{rl} process to gradually adapt to the updated neighborhood statistics.
This approach enables RL-ASL to preserve connectivity and progressively re-optimize performance under moderate mobility without requiring major protocol modifications.

\subsection{Routing Layer Interaction and Adaptation}

RL-ASL operates at the \acrshort{mac} layer and remains agnostic to the routing protocol above it. This modular design eases integration into existing TSCH stacks, as RL-ASL relies only on local per-neighbor statistics (e.g., reception success, inter-arrival intervals) and the node's \acrshort{asn}, which are routing-independent.

Our evaluation focused on static topologies to isolate MAC-layer adaptation effects. However, routing-layer dynamics—such as parent changes or topology updates in RPL—can transiently modify neighborhood relationships. In such cases, RL-ASL automatically reinitializes its \acrshort{2h} sequence and local neighbor statistics, enabling rapid re-synchronization without global reconfiguration.

Future work could strengthen routing-MAC interaction by introducing cross-layer signaling. For example, routing events (e.g., DAO updates or parent switches) could trigger a temporary exploration phase where RL-ASL prioritizes listening and accelerates learning for new neighbors. This would preserve energy efficiency while enhancing robustness in mobile or time-varying networks.

\subsection{Summary}

RL-ASL integrates:
\begin{enumerate*}[label=(\roman*)]
  \item a probabilistic temporal transmission model,
  \item compact state aggregation via mixed-radix encoding, and
  \item expected-reward-driven tabular Q-learning
\end{enumerate*}
to balance energy efficiency and reliability on a per-node basis.
The fully distributed design requires no message exchange or
synchronization across nodes.
Heuristics for missed-neighbor and near-transmit situations improve
convergence stability and robustness under realistic \acrshort{tsch} dynamics.
Algorithm~\ref{alg:rlasl} mirrors the embedded implementation and
is optimized for memory- and timing-constrained IoT devices.

\section{Implementation and Experimental Setup}
\label{sec:implementation}

This section describes the implementation of RL-ASL in \textsc{Contiki-NG}~\cite{oikonomouContikiNGOpenSource2022a}, the experimental environment on the FIT IoT-LAB~\cite{adjihFITIoTLABLarge2015a} testbed, and the configurations used to evaluate its performance under diverse network and traffic conditions.

\subsection{Integration of RL-ASL in Contiki-NG}
\label{sec:implementation_details}

RL-ASL was implemented natively in \textsc{Contiki-NG} (v5.0), an open-source operating system for low-power and lossy networks (LLNs) with built-in support for IEEE~802.15.4, 6LoWPAN, RPL, and the \acrshort{tsch} MAC protocol. This provides a realistic software stack for evaluating learning-based MAC-layer mechanisms.

The RL-ASL module extends the \emph{Orchestra}~\cite{duquennoyOrchestraRobustMesh2015a} scheduling framework by introducing a \acrshort{rl}-driven adaptive listening mechanism. Each node autonomously decides whether to activate or skip its scheduled unicast receive (Rx) slot, reducing idle listening while preserving reliability.

At runtime, neighboring nodes coordinate through a lightweight \emph{\acrshort{2h}} mechanism that synchronizes the transmitter and receiver before each transmission attempt. The handshake succeeds if
\[
    \text{\acrshort{2h}}_{n\to m} =
    \begin{cases}
        1, & \text{if } l^{\mathrm{Tx}}_n = l^{\mathrm{Rx}}_m \land c_n = c_m \land \text{ack}_n = 1, \\
        0, & \text{otherwise,}
    \end{cases}
\]
where \(l^{\mathrm{Tx}}_n\) and \(l^{\mathrm{Rx}}_m\) denote the scheduled slot indices of nodes \(n\) and \(m\), \(c_n, c_m\) their channel offsets, and \(\text{ack}_n\) the acknowledgment flag.
If a node updates its preferred next-hop, the \acrshort{2h} sequence is reinitiated, allowing the previous receiver to safely skip its next Rx slot.

The Q-learning component operates in two phases:
\begin{enumerate}
    \item \textbf{Offline training:} performed entirely in the Cooja network simulator~\cite{osterlindCrossLevelSensorNetwork2006a}, using diverse traffic patterns to update Q-values and converge to an optimal slot-skipping policy.
    \item \textbf{On-device inference:} executed in real hardware using the pre-trained Q-table, stored directly in firmware.
\end{enumerate}

\subsection{Experimental Platform}

All experiments were conducted on the \textbf{FIT IoT-LAB} testbed---a large-scale open-access facility for experimentation with low-power wireless systems. The deployment utilized the \textit{IoT-LAB M3} platform, which integrates an ARM Cortex-M3 microcontroller, an Atmel AT86RF231 IEEE~802.15.4 radio, and onboard sensors for temperature, light, and acceleration. This platform provides a representative low-power embedded system with constrained computational and memory resources, while enabling detailed monitoring of energy, radio, and timing metrics at the hardware level.

Each node runs \textsc{Contiki-NG} with full \acrshort{tsch} synchronization and fine-grained energy instrumentation, allowing precise measurement of duty cycle, latency, and packet reliability. The controlled experimental environment and large number of available nodes allow reproducible and scalable evaluation of RL-ASL under realistic operating conditions.

\begin{figure}[t]
    \centering
    \subfloat[\label{fig:topology-a}]{%
        \resizebox{0.25\columnwidth}{!}{%
            \begin{tikzpicture}[
                    >=latex,
                    every edge/.append style={thick, ->, shorten >=1pt, shorten <=1pt},
                    transform shape=false
                ]

                \pgfmathsetmacro{\SINKSIZE}{10}
                \pgfmathsetmacro{\RELAYSIZE}{10}
                \pgfmathsetmacro{\LEAFSIZE}{10}

                \tikzset{
                    sink/.style={circle, draw=red!70!black, fill=red!20, thick,
                            minimum size=\SINKSIZE pt, inner sep=0pt, font=\bfseries\scriptsize},
                    relay/.style={circle, draw=green!60!black, fill=green!20, thick,
                            minimum size=\RELAYSIZE pt, inner sep=0pt, font=\scriptsize},
                    leaf/.style={circle, draw=blue!60!black, fill=blue!15, thick,
                            minimum size=\LEAFSIZE pt, inner sep=0pt, font=\scriptsize},
                }

                \def\linkscale{0.5}

                \node[sink] (s) at (0,0) {1};
                \node[relay] (r) at (0,-1.6*\linkscale) {2};
                \node[leaf] (l1) at (-0.9*\linkscale,-2.9*\linkscale) {3};
                \node[leaf] (l2) at (0,-2.9*\linkscale) {4};
                \node[leaf] (l3) at (0.9*\linkscale,-2.9*\linkscale) {5};

                \draw (r)--(s);
                \foreach \x in {l1,l2,l3}{\draw (\x)--(r);}
                \begin{scope}[shift={(1.0,0)}]
                    \node[sink,label=right:{\footnotesize Sink}] at (0,1) {};
                    \node[relay,label=right:{\footnotesize Relay}] at (0,0.5) {};
                    \node[leaf,label=right:{\footnotesize Leaf}] at (0,0) {};
                \end{scope}

            \end{tikzpicture}
        }%
    }\hfill
    \subfloat[\label{fig:topology-b}]{%
        \resizebox{0.35\columnwidth}{!}{%
            \begin{tikzpicture}[
                    >=latex,
                    every edge/.append style={thick, ->, shorten >=1pt, shorten <=1pt},
                    transform shape=false
                ]

                \pgfmathsetmacro{\SINKSIZE}{15}
                \pgfmathsetmacro{\RELAYSIZE}{15}
                \pgfmathsetmacro{\LEAFSIZE}{15}

                \tikzset{
                    sink/.style={circle, draw=red!70!black, fill=red!20, thick,
                            minimum size=\SINKSIZE pt, inner sep=0pt, font=\bfseries},
                    relay/.style={circle, draw=green!60!black, fill=green!20, thick,
                            minimum size=\RELAYSIZE pt, inner sep=0pt},
                    leaf/.style={circle, draw=blue!60!black, fill=blue!15, thick,
                            minimum size=\LEAFSIZE pt, inner sep=0pt},
                }

                \def\rA{0.95}
                \def\rB{1.65}
                \def\rC{2.35}

                \node[sink] (n1) at (0,0) {1};
                \node[relay] (n2) at (90:\rA)  {2};
                \node[relay] (n3) at (210:\rA) {3};
                \node[relay] (n4) at (330:\rA) {4};
                \draw (n2)--(n1);
                \draw (n3)--(n1);
                \draw (n4)--(n1);

                \node[relay] (n5)  at (70:\rB)  {5};
                \node[relay] (n6)  at (110:\rB)   {6};
                \node[relay] (n7)  at (190:\rB)  {7};
                \node[relay] (n8)  at (230:\rB)  {8};
                \node[relay] (n9)  at (310:\rB)  {9};
                \node[relay] (n10) at (350:\rB)  {10};
                \foreach \a/\b in {5/2,6/2,7/3,8/3,9/4,10/4}{\draw (n\a)--(n\b);}

                \foreach \a/\ang in {11/62,12/78,13/102,14/118,15/182,16/198,17/222,18/238,19/302,20/318,21/342,22/358}{
                        \node[leaf] (n\a) at (\ang:\rC) {\a};
                    }
                \foreach \a/\b in {11/5,12/5,13/6,14/6,15/7,16/7,17/8,18/8,19/9,20/9,21/10,22/10}{
                        \draw (n\a)--(n\b);
                    }

            \end{tikzpicture}
        }%
    }\hfill
    \subfloat[\label{fig:iotlab-deployment}]{%
        \includegraphics[width=0.4\columnwidth]{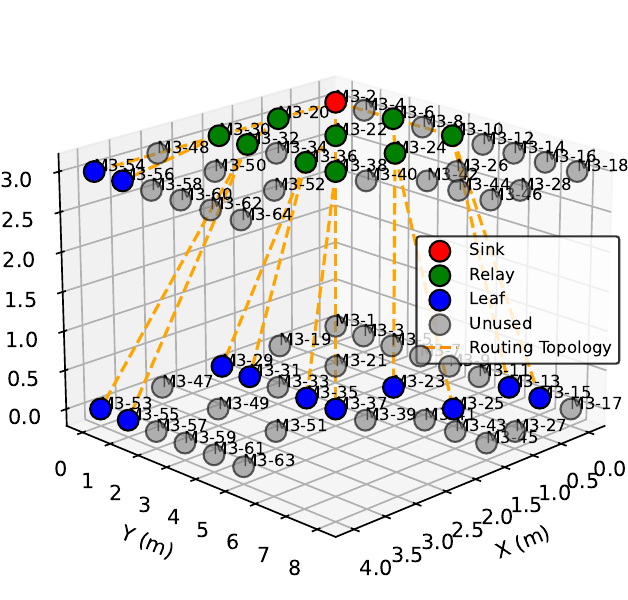}%
    }
    \caption{Experimental network topologies from Strasbourg IoT-LAB site: (a) Simple 5-node topology used for training and validation, (b) Larger star topology, and (c) Deployment of the star topology in the FIT IoT-LAB.}
\end{figure}
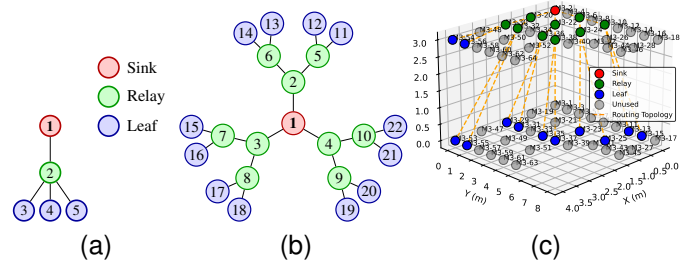

\subsection{Network Scenarios and Topologies}

Two representative network topologies were used to assess RL-ASL:

\begin{itemize}
    \item \textbf{Simple topology:} a compact 5-node network comprising one sink, one relay, and three leaf nodes (Fig.~\ref{fig:topology-a}). It supports controlled validation of per-slot decision dynamics and convergence behavior. \emph{All RL-ASL training was exclusively performed in this topology.}
    \item \textbf{Star topology:} a larger multi-hop deployment derived from the \emph{Strasbourg} IoT-LAB site (Fig.~\ref{fig:topology-b} and \ref{fig:iotlab-deployment}). Leaf nodes are positioned up to three hops from the sink, forming a hybrid star–mesh structure for scalability and latency evaluation under realistic link variability. \emph{This topology only uses the pre-trained, aggregated (\acrshort{fl}) Q-table obtained from the simple topology.}
\end{itemize}

Nodes operate with standard \acrshort{tsch} parameters: 10~ms timeslots, 0~dBm transmission power, and network-wide slot alignment via the \acrshort{asn}.

\subsection{Traffic Patterns}

To evaluate adaptability, multiple traffic generation patterns were implemented as \textsc{Contiki-NG} processes. Each node periodically generates packets aligned with its \acrshort{tsch} schedule.
The bursty traffic behavior evaluated in this work arises from heterogeneous periodic sources rather than from fully random or erratic traffic. In the heterogeneous traffic pattern, nodes generate packets periodically but with different inter-arrival times. When such flows converge at relay nodes, their superposition creates transient congestion and burst-like reception opportunities, especially in multi-hop topologies. This model reflects common industrial monitoring scenarios in which subsets of sensors temporarily report at higher rates due to local events or configuration differences.
Table~\ref{tab:traffic_patterns} summarizes the considered traffic modes and per-node transmission intervals.

\begin{table}[t]
    \centering
    \scriptsize
    \caption{Traffic generation patterns and per-node Tx intervals.}
    \label{tab:traffic_patterns}
    \begin{NiceTabular}[c]{lll}
        \CodeBefore
        \rowcolor{\tableheader}{1}
        \rowcolors{2}{\tablerowcolor}{}[respect-blocks]
        \Body
        \toprule
        \RowStyle{\bfseries}
        Pattern                                     & Jittered             & Transmission intervals per node ID \\
        \midrule
        \textbf{High Traffic}                       & \cmark               & All nodes: 13\,s                   \\
        \Block{4-1}{\textbf{Heterogeneous Traffic}} & \Block{4-1}{\cmark } & IDs 3, 11, 15, 19: 17\,s           \\
                                                    &                      & IDs 4, 12, 16, 20: 30\,s           \\
                                                    &                      & IDs 5, 6, 13, 17, 21: 50\,s        \\
                                                    &                      & IDs 4, 18, 22: 73\,s               \\
        \textbf{Sparse Traffic}                     & \cmark               & Alternating IDs: 60 or 73\,s       \\
        \Block{4-1}{\textbf{Periodic Traffic}}      & \Block{4-1}{\xmark } & IDs 3, 11, 15, 19: 17\,s           \\
                                                    &                      & IDs 4, 12, 16, 20: 19\,s           \\
                                                    &                      & IDs 5, 13, 17, 21: 23\,s           \\
                                                    &                      & IDs 4, 18, 22: 29\,s               \\
        \bottomrule
    \end{NiceTabular}
\end{table}

\subsection{Q-learning Configuration and Hyperparameters}
\label{sec:qlearning_config}

Table~\ref{tab:qlearning_params} summarizes the configuration of the RL-ASL tabular Q-learning agent implemented in \textsc{Contiki-NG}.
The design prioritizes simplicity and computational efficiency for real-time execution on constrained IoT motes.

\begin{table}[t]
    \centering
    \scriptsize
    \caption{RL-ASL Q-learning configuration parameters.}
    \label{tab:qlearning_params}
    \begin{NiceTabular}[c]{ll ll}
        \CodeBefore
        \rowcolor{\tableheader}{1}
        \rowcolors{2}{\tablerowcolor}{}[respect-blocks]
        \Body
        \toprule
        \RowStyle{\bfseries}
        Parameter                                 & Value & Parameter                               & Value \\
        \midrule
        Episode length (\(T_e\))                  & 500   & Discount factor (\(\gamma\))            & 0.9   \\
        Learning rate (\(\alpha\))                & 0.15  & Initial exploration (\(\varepsilon_0\)) & 1.0   \\
        Min. exploration (\(\varepsilon_{\min}\)) & 0.05  & Decay factor (\(\eta_\varepsilon\))     & 0.997 \\
        Reward success (\(R_{\mathrm{succ}}\))    & +1.0  & Skip reward (\(R_{\mathrm{skip}}\))     & +0.5  \\
        Idle cost (\(C_{\mathrm{idle}}\))         & –0.5  & Miss penalty (\(C_{\mathrm{miss}}\))    & –1.0  \\
        Terminal reward (success)                 & +5.0  & Terminal penalty (failure)              & –5.0  \\
        Inter-arrival bins (\(B\))                & 10    & Distance bins (\(D\))                   & 4     \\
        Short-bin threshold (\(b_{\mathrm{th}}\)) & 2     &                                                 \\
        \bottomrule
    \end{NiceTabular}
\end{table}

These parameters yield a discrete state space of
\[
    N_s = 10 \times (3{+}1) \times 4 \times (3{+}1) = 640
\]
states, each associated with two possible actions (listen or skip), resulting in a Q-table of size \(640 \times 2\).
The reward structure promotes energy efficiency by encouraging nodes to skip idle receive slots, while penalizing missed receptions and unnecessary listening.

\subsection{Training and On-Device Inference}

\begin{figure}[t]
    \centering
    \includegraphics[width=0.99\linewidth]{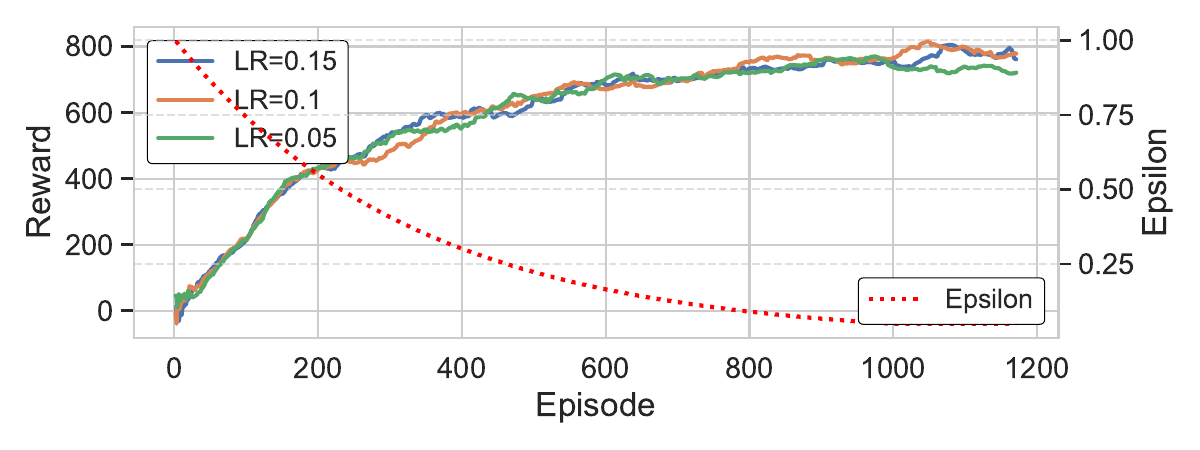}
    \caption{Convergence of average reward during Q-learning training in the relay topology for different learning rates \(\alpha\).}
    \label{fig:rl_training}
\end{figure}

Offline training was conducted exclusively in both the simple topology and the Cooja network simulator.
The Q-learning process is executed entirely offline, and the resulting Q-table is embedded in flash memory as a fixed decision policy, motivated by the constraints of low-power industrial \acrshort{iot} devices.
After deployment, RL-ASL performs no online learning or exploration; runtime behavior is limited to deterministic table lookups based on locally observable state, ensuring predictable execution and compatibility with safety-critical TSCH systems.
Each run simulated $10^8$\,ms (approximately 27.8\,hours) of virtual network time, updating Q-values under an exploration rate $\varepsilon$ decaying from~1.0 to~0.1.
Learning rates $\alpha \in \{0.15, 0.1, 0.05\}$ were evaluated; $\alpha=0.15$ yielded the most stable convergence (Fig.~\ref{fig:rl_training}) and was adopted for deployment.

To enhance generalization across heterogeneous traffic conditions, a lightweight \textit{\acrshort{fl}} aggregation was applied to the Q-tables trained independently for each traffic pattern in the simple topology. Specifically, individual models $\mathbf{Q}_i$ were combined via a weighted \emph{Federated Averaging (FedAvg)} step:
\[
    \mathbf{Q}_{\mathrm{global}} = \sum_i w_i \, \mathbf{Q}_i, \quad
    w_i = \frac{E_i}{\sum_j E_j},
\]
where \(E_i\) denotes the number of episodes used to train model~$i$.
The resulting global model was then deployed directly on real IoT-LAB nodes for inference in both topologies.

The final global Q-table comprises \(640 \times 2 = 1280\) parameters, each stored as a 32-bit floating-point value, resulting in a memory footprint of approximately 5\,kB. Given the limited Flash memory available in low-power motes, this representation can be further optimized through fixed-point quantization. For instance, scaling Q-values by a factor of 10 and storing them as 16-bit integers reduces the footprint to 2.5\,kB with negligible accuracy loss, since the maximum observed Q-value magnitude (\(|Q|_{\max} \approx 33.4\)) comfortably fits within the signed 16-bit range.

The global Q-table was compiled into firmware as a static lookup table. Inference reduces to a table lookup plus two integer comparisons, requiring only a few bytes of RAM for indexing. Flash overhead is platform-dependent: under~10\,kB on the \textit{TI~CC2650} (128\,kB Flash / 20\,kB RAM) and about 5--6\,kB on the \textit{IoT-LAB M3} (STM32F103, 512\,kB Flash / 64\,kB RAM), comfortably within device constraints.

\subsubsection{Energy Consumption of Training}

All training of RL-ASL was conducted within the \textbf{Cooja network simulator}, which provides cycle-accurate emulation of \textsc{Contiki-NG} nodes with realistic radio behavior and timing, ensuring reproducibility without real energy expenditure.
Training directly on embedded hardware is infeasible for low-power IoT networks, as \acrshort{rl} typically requires thousands or millions of interaction steps to converge—resulting in excessive training time and energy use from repeated transmissions and updates. In contrast, simulation in the Cooja network simulator enables accelerated training under controlled yet realistic conditions, faithfully modeling network dynamics and interference.
Once converged, only the compact pre-trained Q-table is deployed on the motes. During operation, inference consists solely of a table lookup, incurring negligible computational and energy cost, as demonstrated in Section~\ref{sec:results}.

\begin{table}[t]
    \centering
    \scriptsize
    \caption{\hl{Sensitivity of RL-ASL to the reward parameter $R_{\mathrm{skip}}$ in the simple topology.}}
    \label{tab:sensitivity}
    \begin{NiceTabular}[c]{lccc}
        \CodeBefore
        \rowcolor{\tableheader}{1}
        \rowcolors{2}{\tablerowcolor}{}[respect-blocks]
        \Body
        \toprule
        \RowStyle{\bfseries}
        $\hl{R_{\mathrm{skip}}}$ & \hl{Latency [ms]} & \hl{PDR [\%]} & \hl{RDC [\%]} \\
        \midrule
        \hl{0.25}                & \hl{202.37}       & \hl{99.91}    & \hl{1.35}     \\
        \hl{0.5}                 & \hl{196.38}       & \hl{99.97}    & \hl{1.32}     \\
        \hl{0.75}                & \hl{198.70}       & \hl{99.97}    & \hl{1.33 }    \\
        \bottomrule
    \end{NiceTabular}
\end{table}

\subsection{Sensitivity Analysis}

\hl{Table~\ref{tab:sensitivity}} illustrates the sensitivity of RL-ASL to the reward parameter $R_{\mathrm{skip}}$ \hl{in the simple topology}.
Across the tested range, all three performance metrics exhibit only minor variations.
In particular, \hl{the} \acrshort{pdr} remains consistently above 99\%, \hl{varying between 99.91\% and 99.97\%.}
\hl{The end-to-end latency ranges from 196.38\,ms to 202.37\,ms, corresponding to a difference of about 6\,ms across all configurations, while the \acrshort{rdc} remains low and stable between 1.32\% and 1.35\%.}

These results indicate that RL-ASL is largely insensitive to moderate changes in reward weighting and does not rely on carefully tuned values of $R_{\mathrm{skip}}$ to maintain reliable and energy-efficient operation within a given deployment scenario.
This insensitivity reduces the need for precise reward calibration during offline training and supports the practical deployability of RL-ASL.

\section{Performance Metrics and Baseline Protocols}
\label{sec:performance_metrics}

This section defines the performance metrics used to evaluate \acrshort{rlasl} and describes the baseline protocols employed for comparison. The evaluation focuses on quantifying the protocol’s ability to improve reliability, latency, and energy efficiency under varying traffic and topology conditions.

\subsection{Performance Metrics}

We consider three key performance metrics to assess \acrshort{rlasl}: \acrshort{pdr}, latency, and power consumption. Together, these metrics provide a comprehensive view of the trade-offs between reliability, responsiveness, and energy efficiency.

\subsubsection{PDR}
The \acrshort{pdr} quantifies link reliability and is defined as the ratio of successfully received packets to the total packets transmitted. For a unicast link from node \( n \) to node \( m \),
\(
\text{PDR}_{n \to m} = \frac{|\Omega_{n \to m}|}{|\Psi_{n \to m}|},
\)
where \( \Omega_{n \to m} \) and \( \Psi_{n \to m} \) denote the sets of successfully received and transmitted packets, respectively.  
The network-wide PDR is computed as the average over all leaf nodes:
\begin{equation}
    \text{PDR} = \frac{\sum_{n \in \mathcal{Z}} |\Omega_{n \to \mathcal{G}}|}{\sum_{n \in \mathcal{Z}} |\Psi_{n \to \mathcal{G}}|},
\end{equation}
where \( \mathcal{Z} \) is the set of leaf nodes and \( \mathcal{G} \) denotes the sink.

\subsubsection{Latency}
Latency measures the end-to-end delay experienced by packets. For a unicast link \( n \to m \), it is defined as:
\(
\mathcal{D}_{n \to m} = \frac{1}{|\Omega_{n \to m}|} \sum_{p \in \Omega_{n \to m}} (t_{\text{rx}}^p - t_{\text{tx}}^p),
\)
where \( t_{\text{tx}}^p \) and \( t_{\text{rx}}^p \) represent the transmission and reception timestamps of packet \( p \).  
The network-wide average latency is then computed as:
\begin{equation}
    \mathcal{D} = \frac{\sum_{n \in N} \sum_{p \in \Omega_{n \to \mathcal{G}}} (t_{\text{rx}}^p - t_{\text{tx}}^p)}{\sum_{n \in N} |\Omega_{n \to \mathcal{G}}|}.
\end{equation}

\subsubsection{Power and Energy Consumption}
The average current consumption (\( \varkappa \)) of each node is computed using the per-state duty cycle reported by \textsc{Contiki-NG}'s built-in \textit{Energest} module, which tracks time spent in transmit, receive, idle, and low-power states.  
For node \( n \) at time \( t \),
\begin{equation}
    \varkappa_n(t) = \sum_{s \in S} D_{n,s}(t) I_s,
\end{equation}
where \( D_{n,s}(t) \) is the fraction of time in state \( s \) and \( I_s \) is the corresponding current draw.

We use the current consumption values specified in the M3 platform datasheet~\cite{adjihFITIoTLABLarge2015a}:  
\(I_{\text{CPU}} = 14.0\ \mathrm{mA}\), \(I_{\text{LPM}} = 0.014\ \mathrm{mA}\), \(I_{\text{DEEP\_LPM}} = 0.002\ \mathrm{mA}\), \(I_{\text{TX}} = 11.6\ \mathrm{mA}\), and \(I_{\text{RX}} = 12.3\ \mathrm{mA}\), with a supply voltage of \(V = 3.3\ \mathrm{V}\). These parameters are used consistently to compute all energy-related metrics presented in Section~\ref{sec:results}.

\subsection{Baseline Protocols}

To contextualize the performance of \acrshort{rlasl}, we compare it against three representative \acrshort{tsch} scheduling approaches spanning autonomous, link-based, and adaptive paradigms:

\begin{itemize}
    \item \textbf{Orchestra} (Orch.)~\cite{duquennoyOrchestraRobustMesh2015a}: A reference autonomous scheduling framework that assigns timeslots based on node roles. We use the receiver-based Orchestra mode with default parameters, which provides a fair baseline for static listening schedules.
    \item \textbf{Link-based Scheduling} (Orch.-LB)~\cite{elstsEmpiricalSurveyAutonomous2020b,kimALICEAutonomousLinkbased2019}: A deterministic link-centric scheduler that allocates timeslots according to link quality metrics. This baseline highlights the gains of \acrshort{rl} compared to structured yet non-adaptive scheduling.
    \item \textbf{PRIL-M}~\cite{scanzioUltralowPowerGreen2024}: A recent probabilistic adaptive listening protocol that dynamically adjusts reception windows based on traffic. PRIL-M represents the current state of the art in adaptive \acrshort{tsch} listening.
\end{itemize}

We implement RL-ASL as an add-on service that operates on top of the Orchestra and link-based scheduling frameworks—denoted \textit{RL-ASL} and \textit{RL-ASL-LB}. As a non‑invasive layer, the service dynamically skips or activates receive slots according to the learned policy while preserving the underlying timeslot allocation and coordination logic of the base schedulers.

Note that PRIL-M is evaluated only under periodic traffic conditions, as its design relies on piggybacking deterministic next-transmission timing information in packet headers. This mechanism assumes strictly periodic traffic and is therefore not applicable to heterogeneous or non-periodic workloads, for which its underlying assumptions no longer hold.

\section{Results and Discussion}
\label{sec:results}

\begin{figure*}[t!]
    \centering
    \subfloat[\label{fig:topology-a-scenario-1-pdr}]{%
        \includegraphics[width=0.245\linewidth]{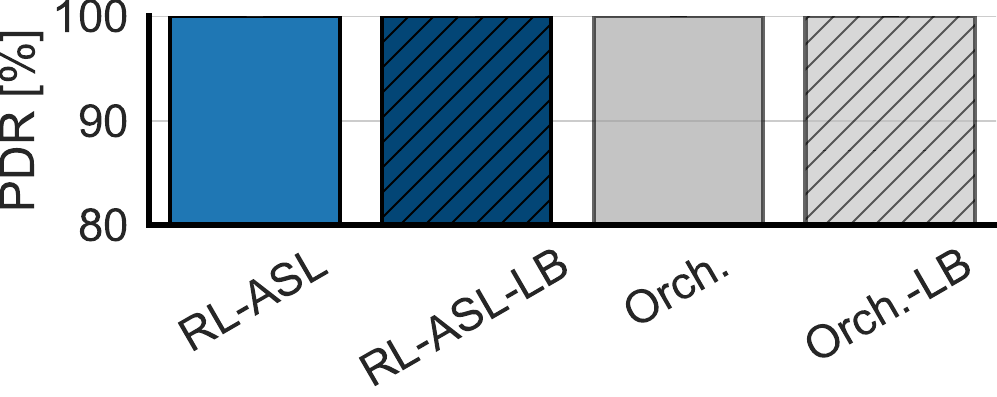}%
    }\hfill%
    \subfloat[\label{fig:topology-a-scenario-2-pdr}]{%
        \includegraphics[width=0.245\linewidth]{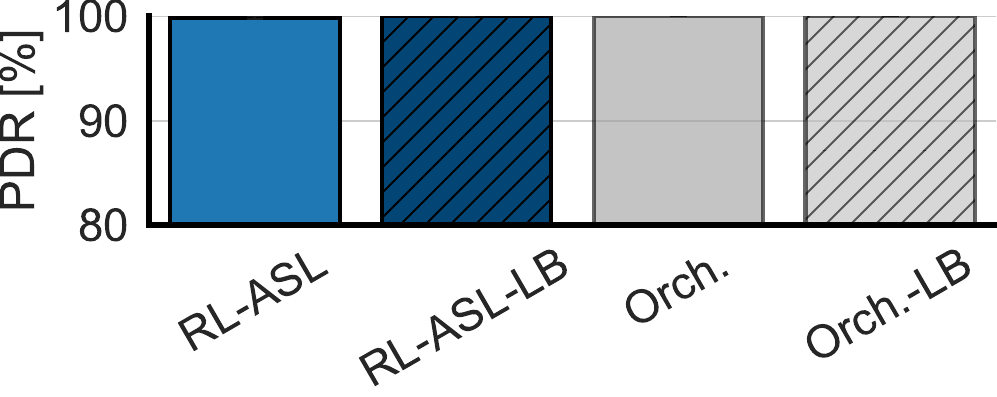}%
    }\hfill%
    \subfloat[\label{fig:topology-a-scenario-3-pdr}]{%
        \includegraphics[width=0.245\linewidth]{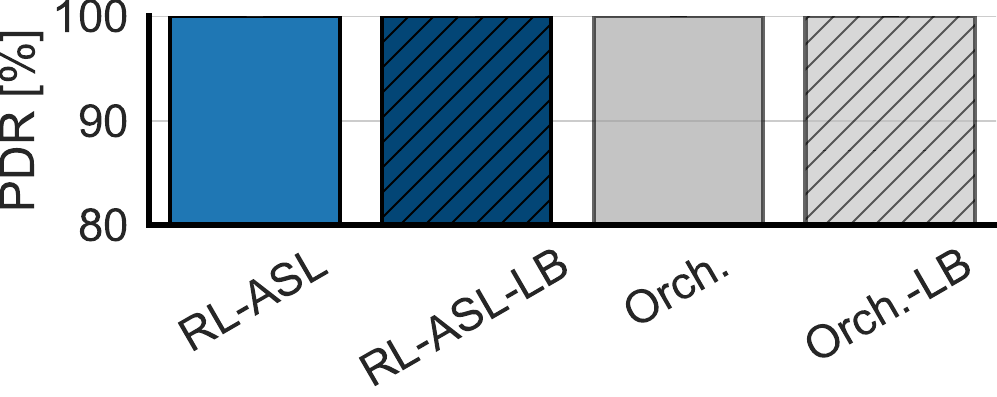}%
    }\hfill%
    \subfloat[\label{fig:topology-a-scenario-4-pdr}]{%
        \includegraphics[width=0.245\linewidth]{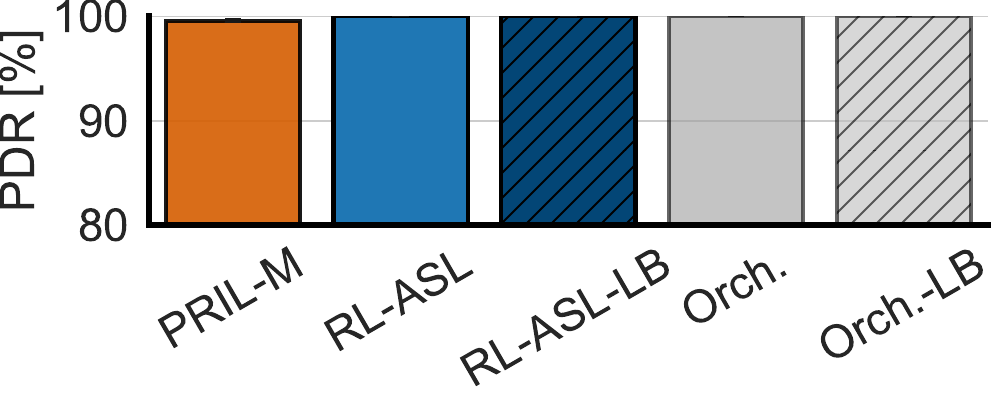}%
    }%
    \caption{Comparison of \acrshort{pdr} across different traffic patterns for the \emph{simple topology}: (a-d) \acrshort{pdr} under high, heterogeneous, sparse, and periodic traffic patterns.}
    \label{fig:results-simple-pdr}
\end{figure*}

\begin{figure*}[t!]
    \centering
    \subfloat[\label{fig:topology-a-scenario-1-latency-violin}]{%
        \includegraphics[width=0.245\linewidth]{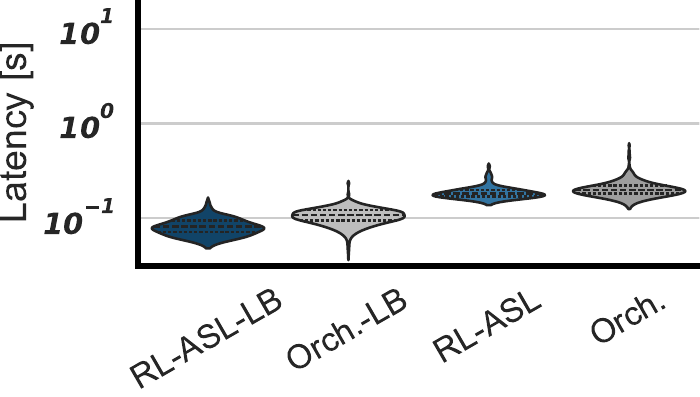}%
    }\hfill%
    \subfloat[\label{fig:topology-a-scenario-2-latency-violin}]{%
        \includegraphics[width=0.245\linewidth]{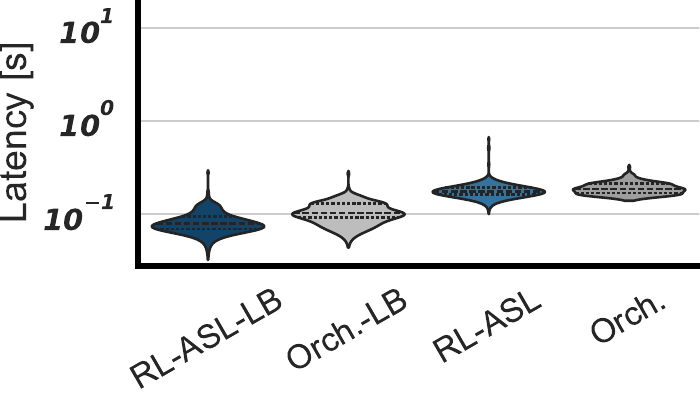}%
    }\hfill%
    \subfloat[\label{fig:topology-a-scenario-3-latency-violin}]{%
        \includegraphics[width=0.245\linewidth]{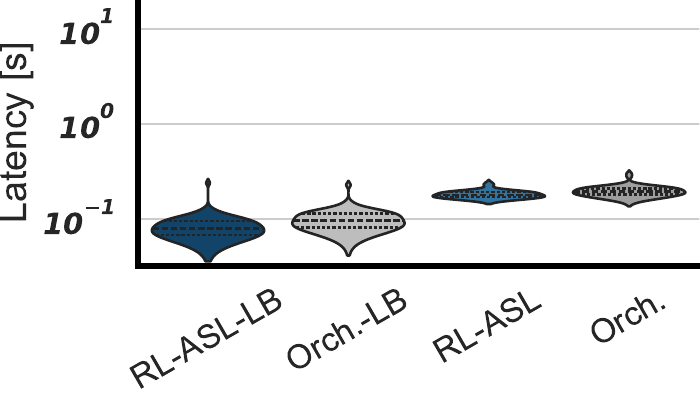}%
    }\hfill%
    \subfloat[\label{fig:topology-a-scenario-4-latency-violin}]{%
        \includegraphics[width=0.245\linewidth]{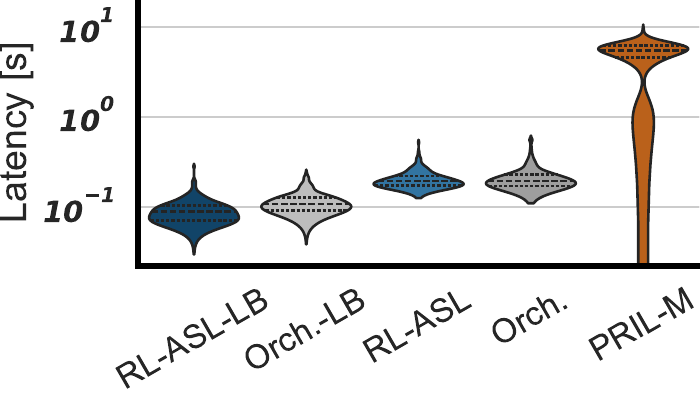}%
    }%
    \caption{Latency distributions across different traffic patterns for the \emph{simple topology}: (a-d) Latency under high, heterogeneous, sparse, and periodic traffic patterns.}
    \label{fig:results-simple-latency}
\end{figure*}

\begin{figure*}[t!]
    \centering
    \subfloat[\label{fig:topology-a-scenario-1-power-stacked_bar}]{%
        \includegraphics[width=0.245\linewidth]{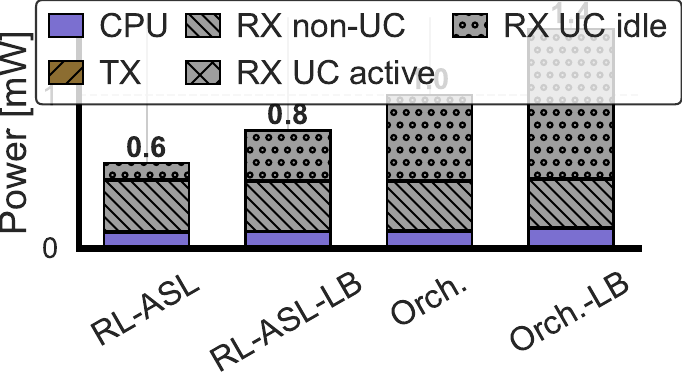}%
    }\hfill%
    \subfloat[\label{fig:topology-a-scenario-2-power-stacked_bar}]{%
        \includegraphics[width=0.245\linewidth]{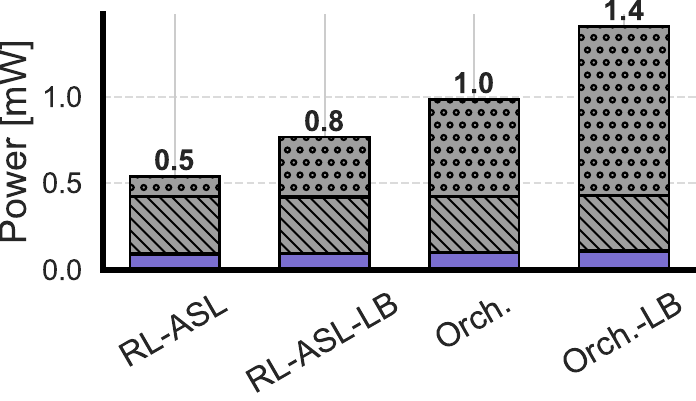}%
    }\hfill%
    \subfloat[\label{fig:topology-a-scenario-3-power-stacked_bar}]{%
        \includegraphics[width=0.245\linewidth]{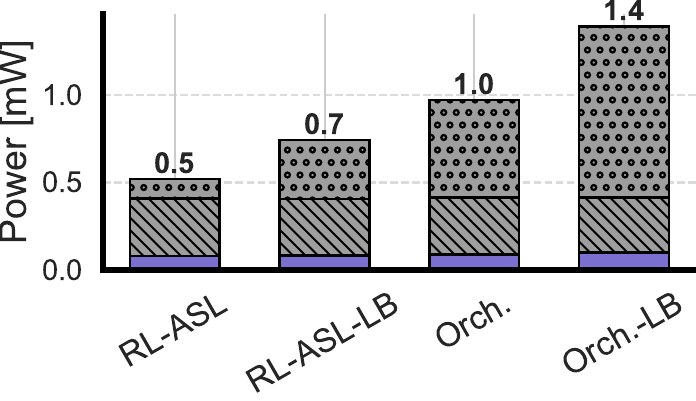}%
    }\hfill%
    \subfloat[\label{fig:topology-a-scenario-4-power-stacked_bar}]{%
        \includegraphics[width=0.245\linewidth]{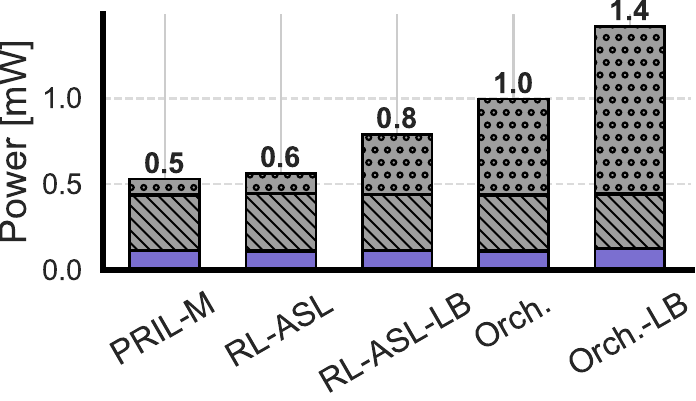}%
    }%
    \caption{Power consumption breakdown across different traffic patterns for the \emph{simple topology}: (a-d) Power consumption under high, heterogeneous, sparse, and periodic traffic patterns.}
    \label{fig:results-simple-power}
\end{figure*}

\begin{figure*}[t!]
    \centering
    \subfloat[\label{fig:topology-a-scenario-1-tradeoff}]{%
        \includegraphics[width=0.245\linewidth]{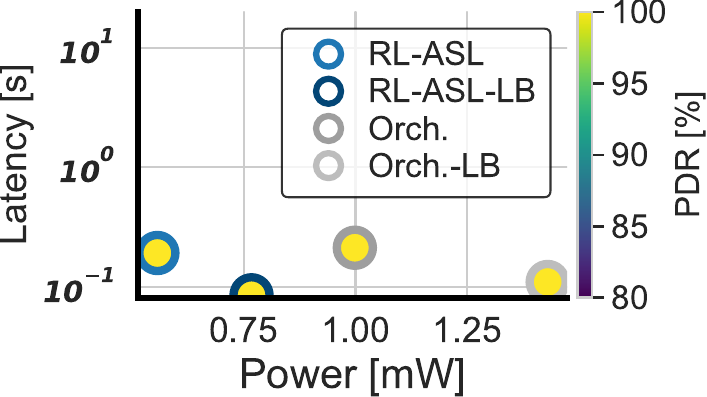}%
    }\hfill%
    \subfloat[\label{fig:topology-a-scenario-2-tradeoff}]{%
        \includegraphics[width=0.245\linewidth]{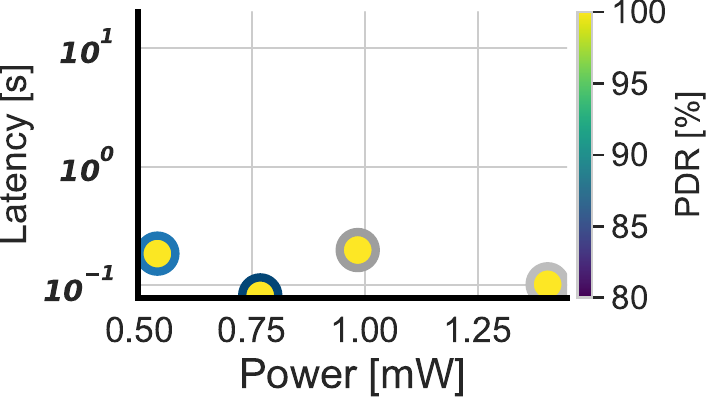}%
    }\hfill%
    \subfloat[\label{fig:topology-a-scenario-3-tradeoff}]{%
        \includegraphics[width=0.245\linewidth]{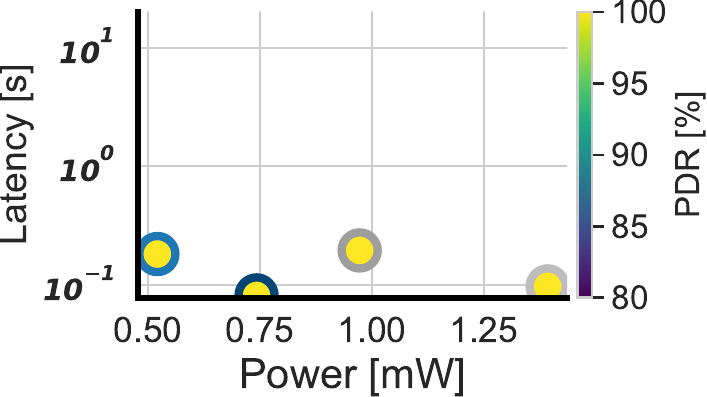}%
    }\hfill%
    \subfloat[\label{fig:topology-a-scenario-4-tradeoff}]{%
        \includegraphics[width=0.245\linewidth]{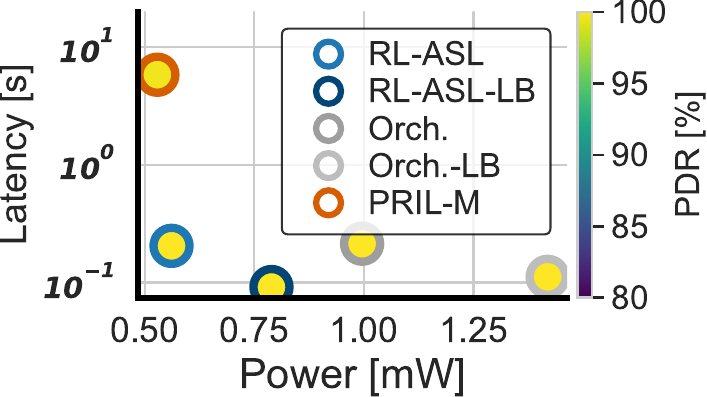}%
    }%
    \caption{Trade-off comparison across different traffic patterns for the \emph{simple topology}: (a-d) Overall performance under high, heterogeneous, sparse, and periodic traffic patterns. The metrics are normalized for comparison.}
    \label{fig:results-simple-tradeoff}
\end{figure*}

Experiments were conducted on the FIT IoT-LAB testbed using two network configurations: a simple linear topology and a complex star topology. For each topology, performance was assessed under four traffic patterns—high, heterogeneous, sparse, and periodic—as described in Section~\ref{sec:implementation}.
It is worth noting that RL-ASL is designed for static and low-mobility TSCH deployments, where network topology remains stable for sufficiently long periods to allow receiver-side listening policies to be learned and applied effectively. Such conditions are typical in many industrial, environmental monitoring, and smart infrastructure scenarios. In highly dynamic networks with continuous or fast mobility, frequent re-synchronization and parent changes may limit the time during which adaptive listening can be exploited.

To ensure statistical robustness, each protocol was executed three times, each lasting 1 hour, and results are reported as the average across all runs.
Error bars in all figures represent 95\% confidence intervals. We analyze results for each topology and traffic pattern, focusing on key performance metrics: \acrshort{pdr}, latency, power consumption, and overall performance (via radar charts).

\subsection{Simple Topology}

We first analyze the results obtained from the simple topology depicted in Fig.~\ref{fig:topology-a}, evaluated under the four traffic patterns. Fig.~\ref{fig:results-simple-pdr}, \ref{fig:results-simple-latency}, \ref{fig:results-simple-power}, and \ref{fig:results-simple-tradeoff} summarize the performance across \acrshort{pdr}, latency, power consumption, and overall trade-offs, respectively.

\subsubsection{Reliability}
Fig.~\ref{fig:topology-a-scenario-1-pdr}-\ref{fig:topology-a-scenario-4-pdr} show that all protocols achieve near-perfect reliability, with network-wide PDRs close to \textbf{100\%}. PRIL-M achieves a slightly lower \acrshort{pdr} (99.5\%) in the periodic scenario.
In contrast, both RL-ASL and RL-ASL-LB maintain a near-perfect \acrshort{pdr} across all traffic patterns, demonstrating that adaptive listening does not compromise reliability. This confirms the robustness of our \acrshort{rl} design, which effectively balances exploration and exploitation while maintaining link stability.

\subsubsection{Latency}
Fig.~\ref{fig:topology-a-scenario-1-latency-violin}-\ref{fig:topology-a-scenario-4-latency-violin} illustrate latency distributions. The main latency differences stem from the underlying \acrshort{tsch} scheduling model. Receiver-based protocols (Orchestra and RL-ASL) exhibit slightly higher delays due to having only one \textit{Rx} cell per slotframe, which increases the average waiting time. In contrast, link-based variants (Orchestra-LB and RL-ASL-LB) achieve lower latency since multiple per-link \textit{Rx} cells reduce queuing delay and enable faster transmissions.

Notably, RL-ASL and RL-ASL-LB achieve latency equal to or slightly lower than their Orchestra counterparts despite employing adaptive listening. This confirms that learning-based slot skipping does not disrupt scheduling continuity. RL-ASL reduces average latency by up to \textbf{96\%} compared to PRIL-M, while RL-ASL-LB achieves up to \textbf{98\%} latency reduction relative to PRIL-M. Overall, RL-ASL maintains low latency even under heterogeneous and sparse traffic, showcasing its adaptability to non-periodic workloads.

\subsubsection{Energy Efficiency}
Fig.~\ref{fig:topology-a-scenario-1-power-stacked_bar}-\ref{fig:topology-a-scenario-4-power-stacked_bar} present the power consumption breakdown across CPU, transmission, and reception states.
RL-ASL reduces total power consumption on average by over \textbf{44\%} compared to Orchestra across all traffic patterns, validating the effectiveness of adaptive listening in minimizing idle-listening energy waste.
Similarly, RL-ASL-LB achieves over \textbf{46\%} lower power consumption than Orchestra-LB, demonstrating that per-link scheduling combined with \acrshort{rl}-driven listening optimization yields substantial energy savings. The most significant savings occur in the \textit{Rx unicast idle} state, as adaptive listening effectively suppresses unnecessary listening intervals while preserving delivery performance.

Importantly, CPU power remains nearly constant across all protocols, indicating that RL-ASL's online inference introduces negligible computational overhead—an essential property for embedded devices. Transmission power also remains comparable, confirming that reduced listening does not incur retransmission overhead. PRIL-M achieves slightly lower power in the periodic case due to its traffic awareness; however, RL-ASL matches this efficiency while supporting arbitrary, non-periodic traffic patterns.

\subsubsection{Overall Performance}
Fig.~\ref{fig:topology-a-scenario-1-tradeoff}-\ref{fig:topology-a-scenario-4-tradeoff} present the overall performance trade-offs involving \acrshort{pdr}, latency, and power consumption.
Both RL-ASL and RL-ASL-LB consistently achieve the best balance across all metrics and traffic patterns.
RL-ASL offers the most energy-efficient operation with slightly higher latency, while RL-ASL-LB provides lower latency at a modest energy cost.
Compared to PRIL-M, RL-ASL delivers comparable power efficiency under periodic traffic but provides lower latency and generalizes effectively to non-periodic and heterogeneous workloads—an essential capability for real-world IoT deployments.

\begin{figure*}[t!]
    \centering
    \subfloat[\label{fig:topology-b-scenario-1-pdr}]{%
        \includegraphics[width=0.245\linewidth]{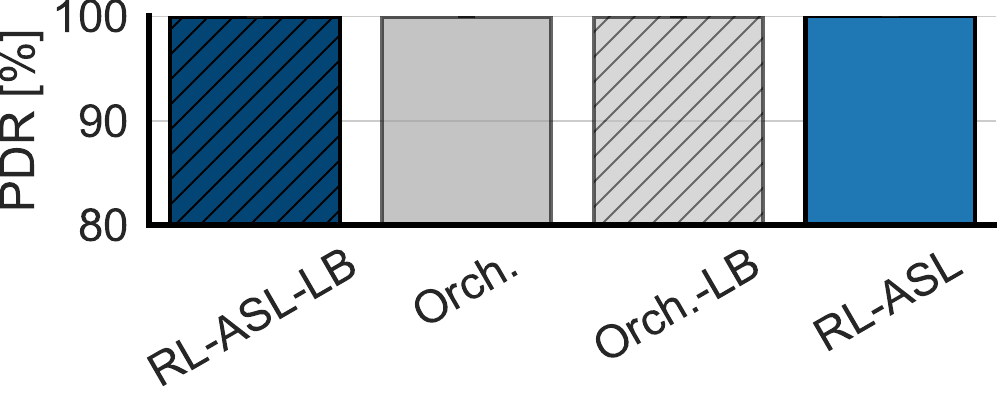}%
    }\hfill%
    \subfloat[\label{fig:topology-b-scenario-2-pdr}]{%
        \includegraphics[width=0.245\linewidth]{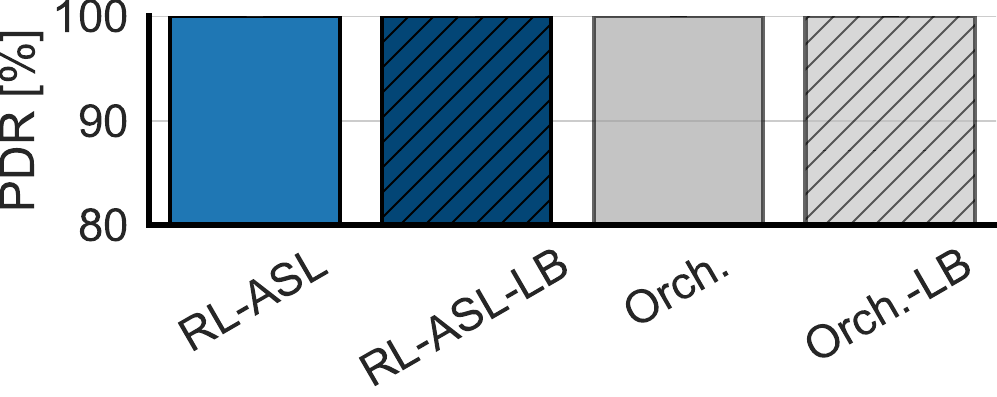}%
    }\hfill%    
    \subfloat[\label{fig:topology-b-scenario-3-pdr}]{%
        \includegraphics[width=0.245\linewidth]{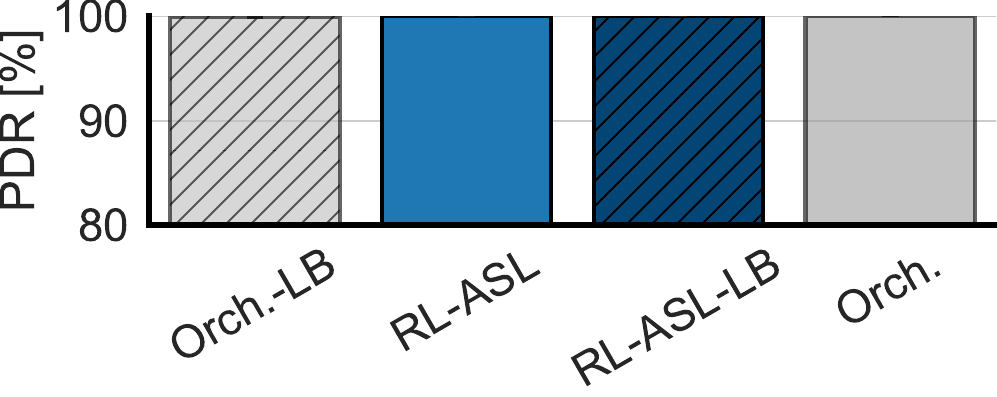}%
    }\hfill%
    \subfloat[\label{fig:topology-b-scenario-4-pdr}]{%
        \includegraphics[width=0.245\linewidth]{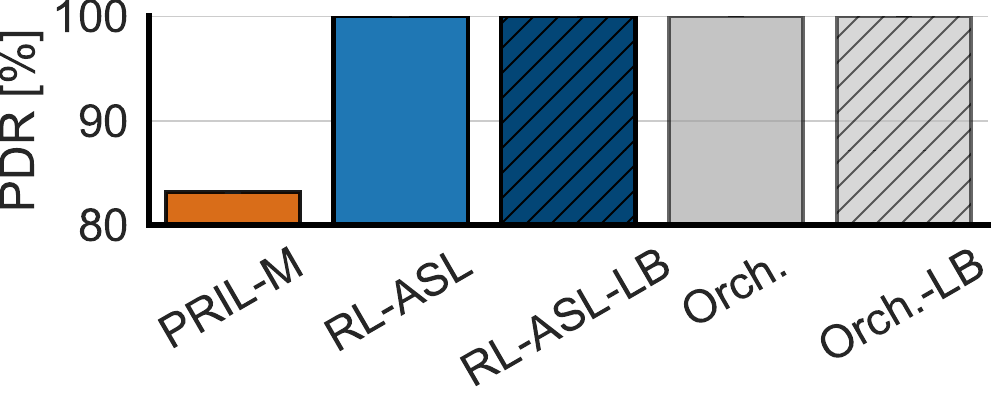}%
    }%
    \caption{Comparison of \acrshort{pdr} across different traffic patterns for the \emph{Star topology}: (a-d) \acrshort{pdr} under high, heterogeneous, sparse, and periodic traffic patterns.}
    \label{fig:results-topology-b-pdr}
\end{figure*}

\begin{figure*}[t!]
    \centering
    \subfloat[\label{fig:topology-b-scenario-1-latency-violin}]{%
        \includegraphics[width=0.245\linewidth]{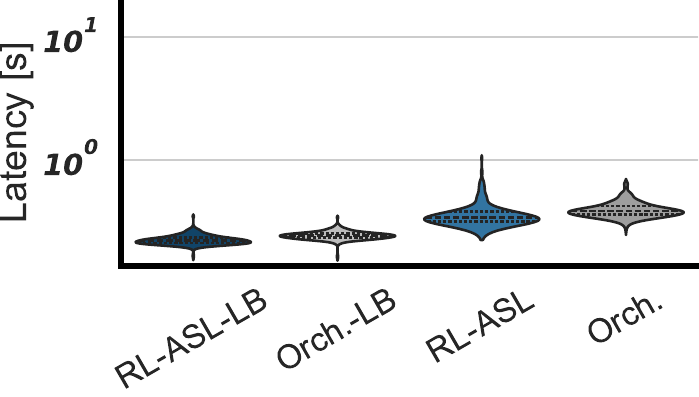}%
    }\hfill%
    \subfloat[\label{fig:topology-b-scenario-2-latency-violin}]{%
        \includegraphics[width=0.245\linewidth]{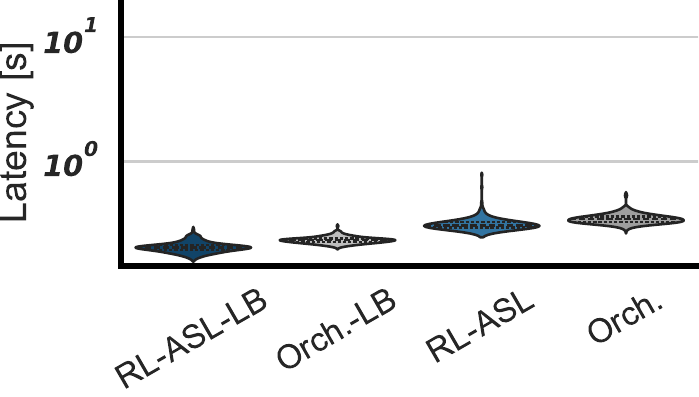}%
    }\hfill%
    \subfloat[\label{fig:topology-b-scenario-3-latency-violin}]{%
        \includegraphics[width=0.245\linewidth]{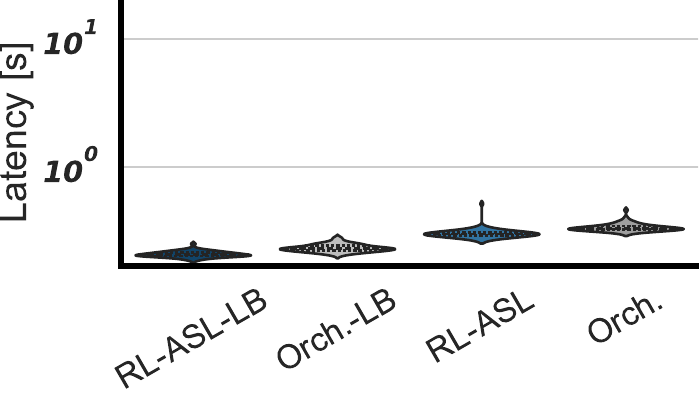}%
    }\hfill%
    \subfloat[\label{fig:topology-b-scenario-4-latency-violin}]{%
        \includegraphics[width=0.245\linewidth]{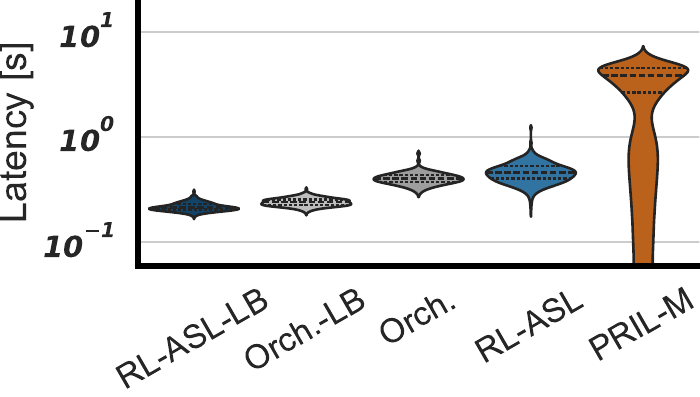}%
    }%
    \caption{Latency distributions across different traffic patterns for the \emph{Star topology}: (a-d) Latency under high, heterogeneous, sparse, and periodic traffic patterns.}
    \label{fig:results-topology-b-latency}
\end{figure*}

\begin{figure*}[t!]
    \centering
    \subfloat[\label{fig:topology-b-scenario-1-power-stacked_bar}]{%
        \includegraphics[width=0.245\linewidth]{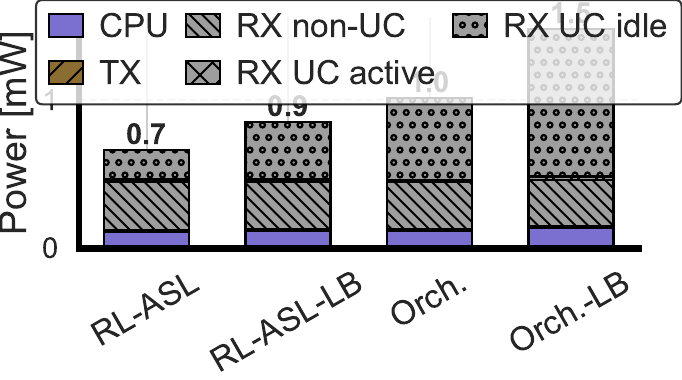}%
    }\hfill%
    \subfloat[\label{fig:topology-b-scenario-2-power-stacked_bar}]{%
        \includegraphics[width=0.245\linewidth]{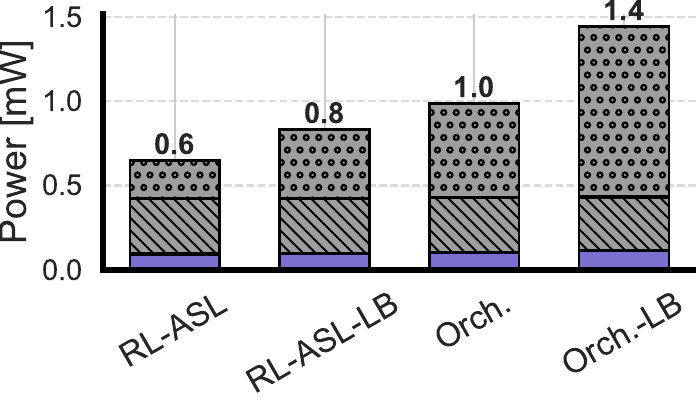}%
    }\hfill%
    \subfloat[\label{fig:topology-b-scenario-3-power-stacked_bar}]{%
        \includegraphics[width=0.245\linewidth]{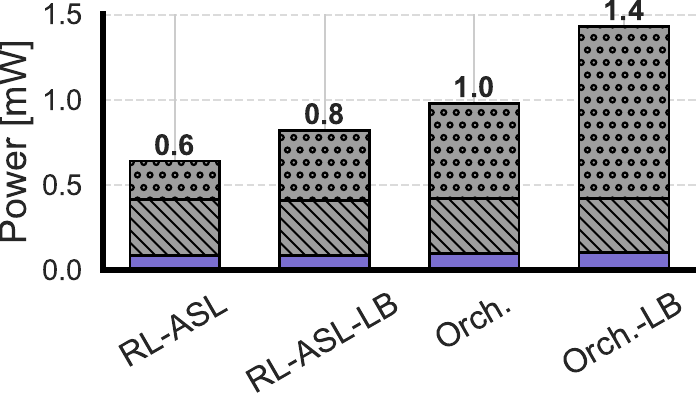}%
    }\hfill%
    \subfloat[\label{fig:topology-b-scenario-4-power-stacked_bar}]{%
        \includegraphics[width=0.245\linewidth]{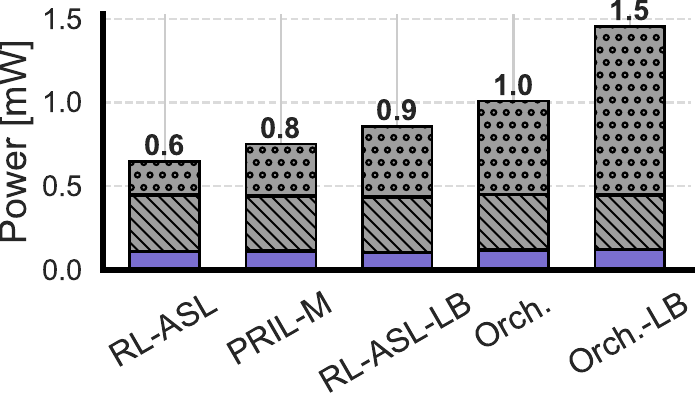}%
    }%
    \caption{Power consumption across different traffic patterns for the \emph{Star topology}: (a-d) Power consumption under high, heterogeneous, sparse, and periodic traffic patterns.}
    \label{fig:results-topology-b-power}
\end{figure*}

\begin{figure*}[t!]
    \centering
    \subfloat[\label{fig:topology-b-scenario-1-tradeoff}]{%
        \includegraphics[width=0.245\linewidth]{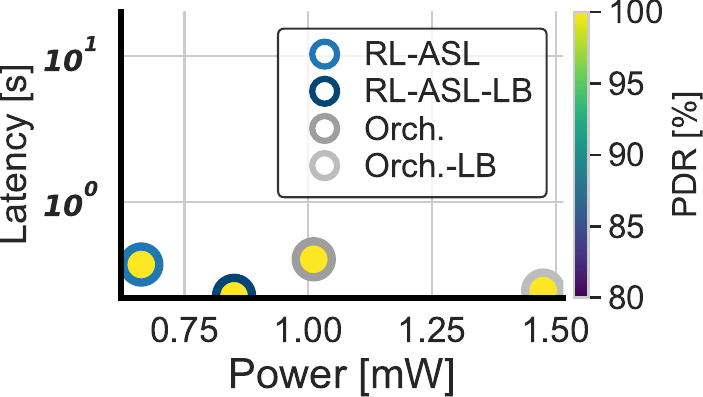}%
    }\hfill%
    \subfloat[\label{fig:topology-b-scenario-2-tradeoff}]{%
        \includegraphics[width=0.245\linewidth]{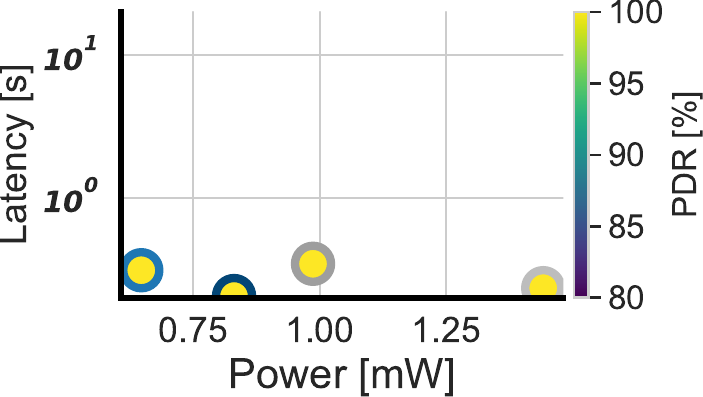}%
    }\hfill%
    \subfloat[\label{fig:topology-b-scenario-3-tradeoff}]{%
        \includegraphics[width=0.245\linewidth]{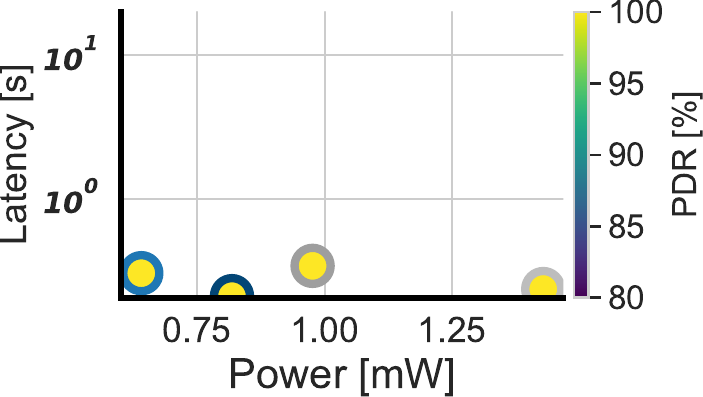}%
    }\hfill%
    \subfloat[\label{fig:topology-b-scenario-4-tradeoff}]{%
        \includegraphics[width=0.245\linewidth]{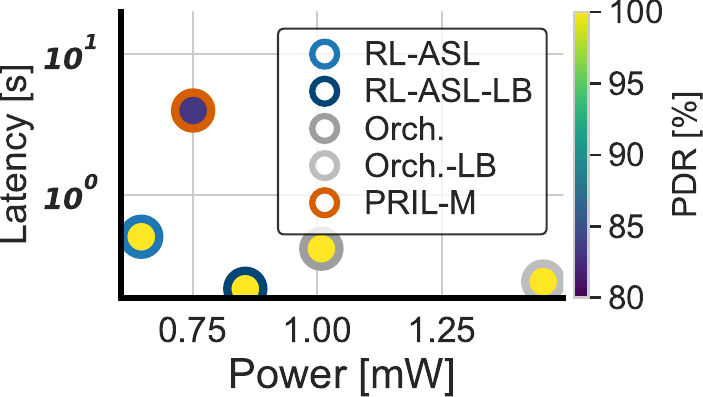}%
    }%
    \caption{Trade-off comparison across different traffic patterns for the \emph{Star topology}: (a-d) Overall performance under high, heterogeneous, sparse, and periodic traffic patterns. The metrics are normalized for comparison.}
    \label{fig:results-topology-b-tradeoff}
\end{figure*}

\subsection{Star Topology}

We next evaluate the star topology (Fig.~\ref{fig:topology-b}), which introduces higher contention and asymmetric traffic. The results, shown in Fig.~\ref{fig:results-topology-b-pdr}-\ref{fig:results-topology-b-tradeoff}, confirm the scalability and robustness of RL-ASL under more complex conditions.

\subsubsection{Reliability and Latency}
Fig.~\ref{fig:topology-b-scenario-1-pdr}-\ref{fig:topology-b-scenario-4-pdr} show that RL-ASL and RL-ASL-LB sustain a 100\% \acrshort{pdr} under all traffic patterns, while PRIL-M's \acrshort{pdr} drops to 83\%.
This demonstrates the robustness of \acrshort{rlasl} in handling contention and asymmetric traffic without sacrificing reliability.

Latency results (Figs.~\ref{fig:topology-b-scenario-1-latency-violin}-\ref{fig:topology-b-scenario-4-latency-violin}) further emphasize the benefits of \acrshort{rl}.
RL-ASL reduces average latency by up to \textbf{87\%} compared to PRIL-M, while RL-ASL-LB achieves reductions of up to \textbf{95\%}.
Latency remains stable despite increased contention in the star topology, indicating that \acrshort{rlasl} effectively manages slot skipping without introducing additional delay.

\subsubsection{Energy Efficiency and Multi-Metric Trade-offs}
Fig.~\ref{fig:topology-b-scenario-1-power-stacked_bar}-\ref{fig:topology-b-scenario-4-power-stacked_bar} show that RL-ASL maintains the lowest total power consumption among all protocols, outperforming PRIL-M even in the periodic traffic scenario.
This highlights that \acrshort{rlasl} adapts to both temporal and spatial dynamics rather than relying on static periodicity assumptions.
Reductions in idle listening power exceed \textbf{35\%} compared to Orchestra and reach up to \textbf{56\%} compared to Orchestra-LB, confirming the effectiveness of adaptive listening in complex topologies.
RL-ASL-LB achieves similar savings, demonstrating that per-link scheduling combined with \acrshort{rl}-based listening optimization remains effective under high contention.

Finally, Fig.~\ref{fig:topology-b-scenario-1-tradeoff}-\ref{fig:topology-b-scenario-4-tradeoff} show that both RL-ASL and RL-ASL-LB consistently achieve the best overall performance across all evaluated metrics and traffic patterns in the star topology.
RL-ASL yields the most energy-efficient operation, at the cost of a slight increase in end-to-end latency, whereas RL-ASL-LB shows lower latency with a modest increase in energy consumption.
In comparison to PRIL-M, RL-ASL demonstrates substantially improved energy efficiency, lower latency, and higher reliability in the periodic traffic scenario—which represents the most favorable operating condition for PRIL-M—while also maintaining robust performance under non-periodic and heterogeneous traffic workloads.
These results indicate that, despite being trained on a simple topology, the learned policy generalizes effectively and adapts to more complex network structures and traffic conditions that were not seen during training.

\subsection{Practical Energy Impact}

To assess the practical benefit of RL-ASL, we estimate the expected battery \acrfull{lt} from the measured average power consumption.
Assuming a $3~\mathrm{V}$ supply and a $220~\mathrm{mAh}$ coin-cell battery ($E_\text{batt} \approx 2.38~\mathrm{kJ}$), the expected \acrshort{lt} in days is:
\[
    L = \frac{E_\text{batt}}{P_\text{avg} \times 86400},
    \quad \text{where } E_\text{batt} = 3 \times 220 \times 3.6 = 2376~\mathrm{J}.
\]
Table~\ref{tab:power-summary} summarizes the measured average power and the corresponding \acrshort{lt} for both network topologies.
In the \textbf{simple topology}, RL-ASL consumes only $0.56~\mathrm{mW}$, extending \acrshort{lt} to about \textbf{174 days}—a \textbf{77\%} gain over Orchestra and \textbf{180\%} over Orchestra-LB.
Even the more responsive RL-ASL-LB maintains \textbf{126 days}, still outperforming static scheduling.
PRIL-M achieves the highest efficiency under strictly periodic traffic ($0.53~\mathrm{mW}$, 184 days), while RL-ASL generalizes better to non-periodic workloads.

In the \textbf{star topology}, RL-ASL ($0.65~\mathrm{mW}$) achieves an estimated \textbf{150 days}, improving \acrshort{lt} by roughly \textbf{55\%} relative to Orchestra ($1.01~\mathrm{mW}$, 97 days) and more than \textbf{120\%} over Orchestra-LB ($1.45~\mathrm{mW}$, 68 days). RL-ASL-LB ($0.86~\mathrm{mW}$) sustains about \textbf{115 days}, balancing adaptivity and energy use.

\begin{table}[t]
    \centering
    \scriptsize
    \caption{Average Power and Estimated Lifetime (3 V, 220 mAh Battery)}
    \label{tab:power-summary}
    \begin{NiceTabular}[c]{lcccc}
        \CodeBefore
        \rowcolor{\tableheader}{1}
        \rowcolors{2}{\tablerowcolor}{}[respect-blocks]
        \Body
        \toprule
        \RowStyle{\bfseries}
        Protocol     & Simple [mW] & LT [days] & Star [mW] & LT [days] \\
        \midrule
        Orchestra    & 0.996       & 99        & 1.008     & 98        \\
        Orchestra-LB & 1.419       & 70        & 1.453     & 68        \\
        PRIL-M       & 0.529       & 185       & 0.751     & 131       \\
        RL-ASL       & 0.561       & 174       & 0.647     & 152       \\
        RL-ASL-LB    & 0.789       & 124       & 0.856     & 114       \\
        \bottomrule
    \end{NiceTabular}
\end{table}

For larger nodes powered by two AA batteries ($\approx21.6~\mathrm{kJ}$), RL-ASL could extend \acrshort{lt} from roughly 2.4 years (Orchestra) to \textbf{4.2 years}, with RL-ASL-LB sustaining around \textbf{3 years}—a notable advantage for long-lived IIoT deployments.

\subsection{Impact of Moderate Mobility on RL-ASL}

\begin{table}[t]
    \centering
    \scriptsize
    \caption{Impact of Moderate Mobility on RL-ASL Performance}
    \label{tab:mobility-impact}
    \begin{NiceTabular}[c]{lccc}
        \CodeBefore
        \rowcolor{\tableheader}{1}
        \rowcolors{2}{\tablerowcolor}{}[respect-blocks]
        \Body
        \toprule
        \RowStyle{\bfseries}
        Protocol     & \acrshort{pdr} [\%] & Radio Duty Cycle [\%] & Latency [ms] \\
        \midrule
        RL-ASL       & 93.1                & 2.8                   & 186.97       \\
        RPL Baseline & 96.7                & 3.4                   & 184.70       \\
        \bottomrule
    \end{NiceTabular}
\end{table}

To assess the behavior of RL-ASL under mobility, we conducted a deliberately simple RPL-based simulation using the Cooja network simulator. The scenario involves a single mobile node moving at a low speed (0.2 m/s) and alternately attaching to one of two potential parents in a tree topology. This experiment uses the heterogeneous traffic pattern described earlier and is intended as a controlled stress test rather than a comprehensive mobility evaluation.

As expected, mobility induces repeated parent changes and receiver re-synchronization phases. During these intervals, nodes temporarily revert to standard TSCH operation, limiting the applicability of receiver-side listening optimization. Under these conditions, the RPL baseline achieves a PDR of approximately 96\%, whereas RL-ASL attains 93\%.

Despite this reduction in delivery ratio, RL-ASL significantly reduces the radio duty cycle from 3.4\% to 2.8\%, corresponding to an energy saving of approximately 18\%. This result indicates that RL-ASL continues to reduce idle listening whenever short periods of topology stability are present, even under moderate mobility.

\subsection{Summary of Findings}
Across all experiments, RL-ASL demonstrates consistent superiority in energy efficiency while maintaining perfect reliability and competitive latency. RL-ASL-LB offers slightly lower delays due to its per-link scheduling structure, while RL-ASL provides the best overall energy-delay trade-off.
Compared with PRIL-M, RL-ASL achieves comparable performance under periodic traffic but generalizes effectively to non-periodic and heterogeneous scenarios—an essential capability for real-world IoT deployments.
A targeted mobility stress test further shows that, while frequent parent changes reduce delivery ratio, RL-ASL continues to achieve substantial idle-listening reductions whenever short periods of topology stability are present.

Overall, these results validate the proposed \acrshort{rlasl} framework as a robust, adaptive, and energy-efficient solution for dynamic \acrshort{tsch} networks, combining the reliability of deterministic scheduling with the flexibility of \acrshort{rl}.

\section{Conclusion}
\label{sec:conclusion}

This work presented \textbf{RL-ASL}, a \acrshort{rl}–driven adaptive listening framework for \acrshort{tsch} networks that enhances energy efficiency without compromising reliability or latency.
By integrating learning-based slot skipping into standard \acrshort{tsch} scheduling, RL-ASL enables nodes to adapt their listening behavior to traffic dynamics, achieving significant reductions in idle-listening power while preserving synchronization and delivery guarantees.
Experimental results from the FIT IoT-LAB testbed and Cooja network simulator show that RL-ASL consistently outperforms state-of-the-art protocols such as Orchestra and PRIL-M across multiple topologies and traffic patterns.
It reduces power consumption by up to \textbf{46\%}, maintains \textbf{near-perfect reliability}, and lowers average latency by up to \textbf{96\%} compared to PRIL-M.
The link-based variant, RL-ASL-LB, further improves delay under contention, confirming the scalability of the proposed learning framework.
Model training is performed entirely in simulation, while inference on real motes is reduced to a simple table lookup with negligible computational and energy overhead.
This makes RL-ASL practical for deployment in low-power embedded networks, bridging the gap between simulation-based learning and real-world operation.
A targeted mobility stress test further indicates that, although frequent parent changes reduce delivery ratio, RL-ASL continues to deliver substantial idle-listening reductions whenever short periods of topology stability are present.
In summary, RL-ASL demonstrates that \acrshort{rl} can be effectively integrated into \acrshort{tsch} scheduling to deliver a robust, adaptive, and energy-aware communication framework for next-generation IoT networks.
Future extensions of RL-ASL will explore tighter integration with mobility-aware routing and scheduling mechanisms, enabling coordinated adaptation of parent selection and receiver listening behavior to further enhance performance in mobile and dynamic environments.

\section*{Acknowledgments}
\label{sec:acknowledgment}

The authors acknowledge the use of AI-based tools for minor language refinements, such as grammar, structure, formatting, and spelling, during manuscript preparation.
All intellectual contributions, technical content, and interpretations are solely those of the authors.

\bibliographystyle{IEEEtran}
\bibliography{RL-ASL}

\begin{IEEEbiography}[{\includegraphics[width=1in,height=1.25in,clip,keepaspectratio]{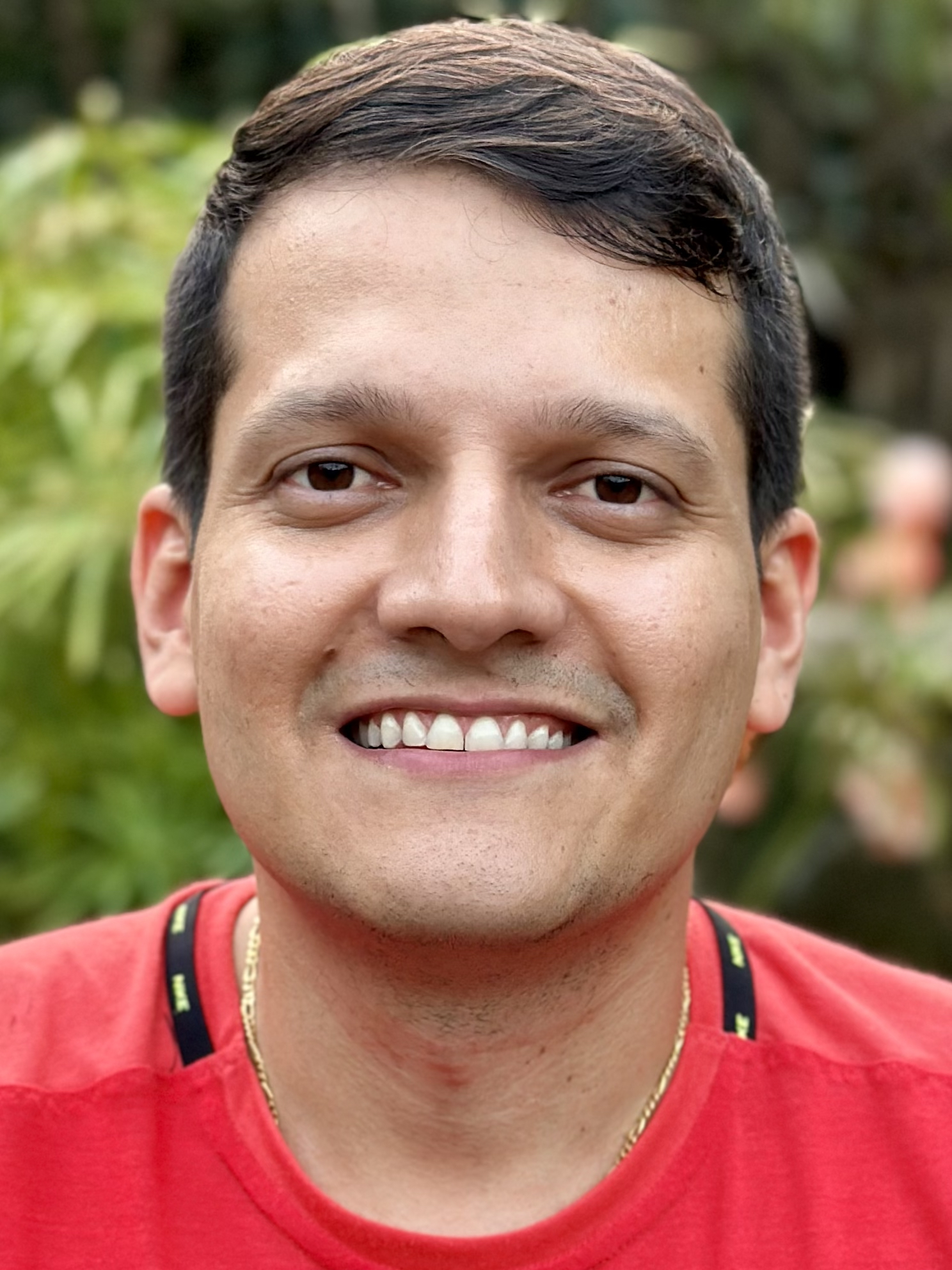}}]{F. Fernando Jurado-Lasso}
  (GS'18--M'21) received the Ph.D. degree in engineering and the M.Eng. degree in telecommunications engineering from The University of Melbourne, Melbourne, VIC, Australia, in 2020 and 2015, respectively, and the B.Eng. degree in electronics engineering from Universidad del Valle, Cali, Colombia, in 2012.

  His research focuses on intelligent networked embedded systems, machine learning for low-power and time-synchronized wireless networks, cross-layer scheduling and resource optimization, and Internet of Things (IoT) protocols and architectures.
\end{IEEEbiography}

\begin{IEEEbiography}[{\includegraphics[width=1in,height=1.25in,clip,keepaspectratio]{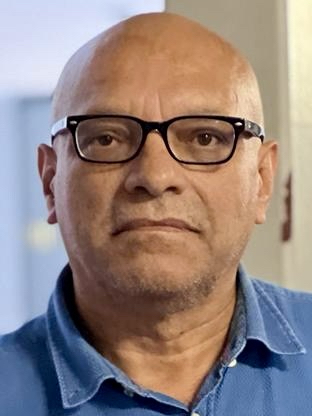}}]{J. F. Jurado}
  received the doctoral degree and M.Sc. degree in physics from Universidad del Valle, Cali, Colombia, in 2000 and 1986, respectively, and the B.Sc. degree in physics from Universidad de Nariño, Pasto, Colombia, in 1984.

  He is currently a Professor with the Department of Basic Sciences, Faculty of Engineering and Administration, Universidad Nacional de Colombia, Palmira, Colombia.
  His research interests include nanomaterials, magnetic and ionic materials, nanoelectronics, embedded systems, and the Internet of Things (IoT).
  He is a Senior Researcher recognized by Minciencias, Colombia, and has been designated as an Emeritus Researcher by Minciencias.
\end{IEEEbiography}

\balance

\vfill

\end{document}